\documentclass[12pt]{iopart}
\usepackage{latexsym}
\usepackage{amssymb}
\begin{document}
\title[Nonlinear vector coherent states
of generalized spin-orbit Hamiltonians]{New classes of nonlinear vector coherent
 states of generalized spin-orbit Hamiltonians}

\author{Joseph Ben Geloun$^{1,2,3}$ and Mahouton Norbert Hounkonnou$^{2}$}

\vspace{10pt}

\address{$^{1}$National Institute for Theoretical Physics NITheP\\
Private Bag X1, Matieland 7602, South Africa\\
$^{2}$International Chair in Mathematical Physics
and Applications \\
(ICMPA--UNESCO Chair) 072 BP 50  Cotonou, Republic of Benin\\
$^{3}$D\'epartement de Math\'ematiques et Informatique\\
 Facult\'e des Sciences et Techniques, Universit\'e Cheikh
Anta Diop, S\'en\'egal}

\vspace{10pt}

\eads{\mailto{bengeloun@sun.ac.za}, \mailto{
norbert.hounkonnou@cipma.uac.bj}\footnote{Correspondence author.} }

\begin{abstract}
This paper deals with an extension of our previous work [J. Phys. A:
Math. Theor. {\bf  40 } F817] by considering an alternative
construction of canonical and deformed vector coherent states (VCSs)
 of the Gazeau-Klauder type associated with generalized spin-orbit Hamiltonians.
We define an annihilation operator which takes into account the finite dimensional
 space of states induced by the $k$-photon transition processes of the two-level atom
 interacting with the single-mode radiation field. The class of nonlinear VCSs (NVCSs) corresponding to the
 action of the annihilation operator is deduced and expressed in terms of generalized displacement operators.
  Various NVCSs  including their ``dual'' counterparts are also discussed. Still by using the Hilbert space structure,
   a new family of NVCSs parameterized by unit vectors  of the $S^3$
  sphere has been identified without making use of the annihilation operator.
\end{abstract}

\pacs{42.50.-p, 03.65.Fd, 02.20.-a}

\vspace{10pt}
 \noindent
\today
\vspace{100pt}

\noindent
NITheP-09-06\\
ICMPA-MPA/2009/11

\maketitle

\section{Introduction}
\label{sect1}

Still arousing the interest for theoreticians,
the vector coherent states (VCSs) \cite{al} have been recently studied
in statistics \cite{ali2} and remain relevant in nonlinear
quantum optics \cite{matos}-\cite{BGH2}.
In particular, their appearance in quantum deformations of physical systems such as
generalized spin-orbit Hamiltonians has been proved in \cite{BGH1}\cite{BGH2}.

Let us precise here the meaning of ``generalized'' spin-orbit
interactions according to \cite{BGH2}. In the context of
semiconductor physics and spintronics, Rashba \cite{rash} and
Dresselhaus \cite{dres} interactions are typical examples of
spin-orbit potentials (for a review of spintronics and spin-Hall
effect, see for instance \cite{js}) which can be recast as
\cite{shen}
\begin{eqnarray}
V_{R}=\,i\,\mu\,(\,a\, \sigma_{-}\,-\,a^{\,\dag} \sigma_{+}),\qquad\quad
V_{D}=\,\lambda\,(\,a\, \sigma_{+}\,+\,a^{\,\dag} \sigma_{-})
\label{eq:vrvd}
\end{eqnarray}
with coupling constants $\mu$ and $\lambda$, respectively;
$a$ and $a^\dag$ are the usual Heisenberg generators
and, given Pauli-spin matrices $\sigma_{1}$,$\sigma_{2}$ and $\sigma_{3}$, we
define $\sigma_{\pm}:=(\sigma_{1}\,\pm i\,\sigma_{2})/2$.

In quantum optics, the Jaynes-Cummings model \cite{jc}, idealizing
the interaction radiation-matter, possesses, in the rotating wave
approximation, a spin-orbit interaction which can be written in the
same form as $V_{D}$ \cite{hus,hus4}.

A first nonlinear spin-orbit interaction, but still remaining in
canonical quantum formulation, is provided by considering a
$k$-photon contribution to an intensity dependent coupling
$\lambda(N)$, a complex function of the operator $N=a^\dag a$. The
introduction of such a number dependent and $k$-multiphoton coupling
becomes significant in the study of the intensity dependent
interaction between a single atom and the radiation field with the
atom making $k$-photon transitions \cite{buck,crnu,agar,nad} as well
as in the study of the quantized motion of a single ion in an
anharmonic oscillator potential trap \cite{sharma,daou}. A second
stage is reached by adding further nonlinearity by the introduction
of $f$-deformed quantum algebras \cite{jannus} defined by the
modified Heisenberg generators coupled to a free continuous function
$f$ of the number operator $N$ such that
\begin{eqnarray}
&& A^{-}=\,a\,f(N),\; A^{+}=\,f(N)\,a^{\dag},
\:\; \{N\}=A^{+} A^{-}=N f^{\,2}(N),\\
&& [A^{-},A^{+}]= \{N+1\}-\{N\}.
\label{eq:f-def}
\end{eqnarray}
Only the limit $f(N)\to 1$ reproduces the ordinary Heisenberg algebra.
It was in this framework introduced by Jannussis {\it et al.} \cite{jannus} that
the earlier notion of ``nonlinear coherent states" (NCSs)
was highlighted. NCSs built from a realistic physical model
are due to de Matos Filho and Vogel \cite{matos} and involve nonclassical properties
and quantum interference effects.
Afterwards, Man'ko and co-workers \cite{manko} interpreted that the $f$-oscillator
action is provided as corresponding to a specific vibration for which the frequency
of oscillation becomes energy dependent.
Still recently, many developments involving the algebra (\ref{eq:f-def})
have been met in special function theory, quantum groups and generalized coherent state (CS)
quantization (see \cite{burban,nk} and references therein).

A $f$-deformed quantum version of $V_{R,D}$ (\ref{eq:vrvd}),
incorporating the heretofore nonlinearities, can be written as
\begin{equation}
\fl  \label{eq:sofke}
\Pi_{k,\varepsilon}=
\, {\mathcal B}^{\,+}_{k,\,\varepsilon}\,\,\sigma_{+}\,+\,
{\mathcal B}^{\,-}_{k,\,\varepsilon}\,\,\sigma_{-},
\qquad
{\mathcal B}_{\,k,\,\varepsilon}^{\,+}:=
A^{-\varepsilon\,k}\,\lambda^{\varepsilon}(N),
\qquad {\mathcal B}_{\,k,\,\varepsilon}^{\,-}:=
\lambda^{-\varepsilon}(N)\,A^{\varepsilon\,k},
\end{equation}
where the symbol $\varepsilon=\pm$ fixes the notations as
$\lambda^{+}(N):=\lambda(N)$ and $\lambda^{-}(N):=\overline{\lambda(N)}$, the bar denoting complex conjugation.

The canonical spin-orbit Hamiltonian includes a dynamical part of
a photon field of frequency $\omega$ and a Zeeman spin term of atomic frequency $\omega_0$ \cite{shen}.
It finds henceforth a nonlinear extension regarding deformed spin-orbit Hamiltonians of the reduced (dimensionless)
form \cite{BGH2}
\begin{eqnarray}
\fl {\mathcal H}^{\rm red}_{k,\varepsilon}
=\frac{(1+\epsilon)}{2}(\{N+1\}+ \{N\})
+\frac{1}{2}(\{N+1\}-\kappa\{N\})\,\sigma_{3} +
\, {\mathcal B}^{\,+}_{k,\,\varepsilon}\,\sigma_{+}\,+\,
{\mathcal B}^{\,-}_{k,\,\varepsilon}\,\sigma_{-}.
\label{eq:fkeqm}
\end{eqnarray}
The ratio $(1+\epsilon)=\omega /\omega_{0}$ determines the rotating wave approximation
if and only if the detuning parameter $\epsilon$ satisfies
$|\epsilon|\ll 1$ and $|\omega-\omega_{0}|\ll \omega,\,\omega_{0}$.
The real parameter $\kappa$ is introduced in order
to recover some known models. A list of significant reduced models
connected to (\ref{eq:fkeqm}),  its canonical limit $f(N)\to 1$
and $\kappa\to 1$ and applications in quantum optics, in condensed matter physics and in semiconductor physics,
  in particular, in the so-called domain of spintronics studying some new spin-dependent phenomena in order to
build a new generation of electronic devices, are available in \cite{BGH2} and references therein.

We shall refer to the model defined by ${\mathcal H}^{\rm
red}_{k,\varepsilon}$  as the $(k,\varepsilon,\kappa,f)$-model, or
more simply, to as the $f$-deformed model. This model has an exactly
solvable spectrum and its Hilbert space of eigenstates can be
decomposed in a finite sequence of $k$ initial states related to
$k$-photon processes and two infinite sequences of states commonly
named ``towers".

In former studies \cite{BGH1,BGH2}, classes of VCSs and nonlinear VCSs (NVCSs) have been
defined for spin-orbit Hamiltonians associated with a nontrivial action
of the annihilation operator only on the two towers. Furthermore, these families of NVCSs
 meet all requirements of Gazeau-Klauder \cite{gk,kl2} with the main vector character formulated
in terms of the unit sphere $S^2$ vectors \cite{BGH1} or labeled by two spin states \cite{BGH2}.
 The case of annihilation operator matrix eigenvalue problem
(with diagonal and quaternion matrices) for NVCSs were successfully treated.
However, these investigations have let aside the initial states induced
by the $k$-transition processes, the annihilation operator canceling
them by definition. It could be then interesting to ask if NVCSs  may include
the finite dimensional space spanned by this limited sequence of states.
Moreover, as far as we can establish, deformed displacement operators
which could generate NVCSs have not been addressed in prior developments. Recalling that
displacement operators for VCSs over matrix domains has been defined by Ali {\it et al} \cite{al}
and also for NCSs (\cite{roy,rok,rok2} and references therein),
 one could investigate their form in the case of NVCSs.

In this paper, we give a new construction of NVCSs (canonical
included) of the Gazeau-Klauder type associated with generalized
spin-orbit Hamiltonians. We define an annihilation operator which
takes into account the finite dimensional space of states of the
initial $k$-photon processes. The class of NVCSs corresponding to
the action of the annihilation operator is expressed in terms of generalized
displacement operators. Issues concerning ``dual" VCSs and $T$
operators \cite{rok2} are tackled and exactly solved for certain
families of parameters. Besides, a new family of NVCSs parameterized
by $S^3$ unit vectors with an exact resolution of the identity has
been identified without making use of the annihilation operator. The latter
class involves the definition of CSs identified onto finite
dimensional Hilbert space \cite{kuang,leons}.

In summary, this paper addresses the following new results:
\begin{enumerate}
\item A generalization of previous constructions of Gazeau-Klauder $S^2$ and matrix
NVCSs for spin-orbit Hamiltonians to  Gazeau-Klauder type $S^2$, normal and  quaternionic NVCSs by including the
finite sequence of states induced by the $k-$photon processes in the definition of
the generalized deformed Barut-Girardello eigenvalue problem. This allows to solve
properly the issue of discontinuity of NVCSs previously observed at $z=0$ in \cite{hus,BGH1,BGH2}.
\item The explicit identification of novel classes of solvable NVCSs (canonical, generalized
 $(p, q)-$deformed in the sense of  \cite{burban,nk}) with exact resolution of the identity characterized by
a new set of deformation parameters covering all related previous results.
 \item An extension to normal matrix domain of the NVCSs with a nontrivial resolution of the identity
 requiring an integration over a $U(2)$ group. This provides a concrete  realization of the formulation
 by Ali {\it et al} \cite{al} in nonlinear deformed models.
 \item The introduction of the concept of deformed displacement operators, deformed dual states
 (with a peculiar temporal stability)  and deformed $T-$operators
 for NVCSs and matrix NVCSs.
 \item The identification of a large class of NVCSs and their dual counterparts (to be classified)
  on the basis of the operator ordering occurring in the construction.
  \item The determination of a new class of $S^3$ NVCSs still defined in the full Hamiltonian Hilbert space by
  assigning a new angle to the $k-$first states.
\end{enumerate}

The paper's outline is as follows. Section 2 recalls in brief considerations
and the Hilbert space structure for classes of spin-orbit models.
A quick review of the NVCSs associated with this Hilbert space is also supplied. Section 3
is devoted to the Hilbert space reorganization and to the definition of ladder operators.
 Sections 4 to 6 deal with
the construction of a family of Gazeau-Klauder NVCSs defined by the action of the annihilation operator
and constituting a generalization of anterior NVCSs of spin-orbit models.
 Generalized displacement operators, dual states and associated $T$ operators are also treated therein.
A different family of $S^3$ NVCSs is investigated in Section 7.
A conclusion is given in Section 8
 and an appendix yields useful relations concerning $(p,q)-$ deformed exponential functions.

\section{Nonlinear vector coherent states of spin-orbit models: a quick review}
\label{sect2}

In this section, a rapid overview of previous results on canonical
and NVCSs drawn from \cite{BGH1,BGH2} is discussed. We aim at
generalizing these results in the remaining sections.

As matter of clarity, let us briefly recall that the bosonic algebra with parameter
$\varepsilon=\pm$ is generated by the Heisenberg operators
$a^{-}:=a$ and $a^{+}:=a^{\,\dag}$ so that $a^{\,\varepsilon}$ is
well defined and $a^{\,-\varepsilon}$ denotes its adjoint. The
Heisenberg-Fock algebra then reads off
$[a^\varepsilon,a^{-\varepsilon}]=-\varepsilon$. For any $k\in
\mathbb{N}$, one defines $a^{\,\varepsilon\,
k}:=(a^{\,\varepsilon})^{\,k}$. These operators acts on the Fock
representation space $F=\{\,|n\rangle,\,\,n\,\in\,\mathbb{N}\,\}$ in
the usual manner. We have, by simple recurrence, for any $k\in
\mathbb{N}$,
\begin{eqnarray}
\label{eq:aeps}
\fl a^{\,\varepsilon\,k}\,|n\rangle=
\left\{\begin{array}{ll}
\left(\frac{(n+\varepsilon k)!}{n!}\right)^{\,\varepsilon/2}|n+\varepsilon\,k\rangle,
& \; {\rm if}\; \;\varepsilon=+,\; n\geq 0 \;
\;{\rm or}\;\;  \varepsilon=-,\;\; n\geq k\,   \\
0, & \; {\rm if}\;\;\varepsilon=-,\; n< k\,
\end{array}\right.
\end{eqnarray}
Given a function $g=g(N)$, by action on the representation space,
the following $\varepsilon$-commutation rules hold
\begin{eqnarray}
a^{\,\varepsilon}\,g(N)\,
=\,g(N-\varepsilon)\, a^{\,\varepsilon},
\qquad
a^{\,\varepsilon\,k}\,g(N)\,=\,g(N-\varepsilon\,k)\,
a^{\,\varepsilon\,k}, \;\; k\in \mathbb{N}.
\label{eq:ecom}
\end{eqnarray}
Let $f=f(N)$ be a fixed nonvanishing operator. Then, we define
$A^{-}:=a^{-}\,f(N)$ and $A^{+}:=\,f(N)\,a^{+}$. Hence the notations
such that $A^{\,\varepsilon}$ and, for $k\in \mathbb{N}$,
$A^{\,\varepsilon\, k}:= (A^{\,\varepsilon})^{\,k}$ make a sense.
One expands $A^{\,\varepsilon\, k}$ as
\begin{eqnarray}
A^{\,\varepsilon\, k}=\left(\frac{f(N)!}{f(N-\varepsilon k)!}
\right)^{\,\varepsilon}\,a^{\,\varepsilon\, k}
=a^{\,\varepsilon\, k}\,\left(\frac{f(N+\varepsilon k)!}{f(N)!}
\right)^{\,\varepsilon},
\end{eqnarray}
where the formal operator $f(N)!$ acts by representation as
$f(N)!\,|n\rangle=f(n)!\,|n\rangle$
with $f(n)!:=f(n)(f(n-1)!)$ and by convention $f(0)!=1$.
Similarly to (\ref{eq:ecom}), the identity
$A^{\,\varepsilon\,k}\,g(N)=g(N-\varepsilon\,k)A^{\,\varepsilon\,k}$ is true.

The $f$-deformed oscillator algebra of Jannussis {\it et al} \cite{jannus} defined by
\begin{eqnarray}
&&A^{-}=a\,f(N),\quad A^{+}=f(N)\,a^{\,\dag},\quad \{N\}:=A^{+}A^{-}=N\,f^{\,2}(N),\cr
&&[A^{-}, A^{+}]= (N+1)\,f^{\,2}(N+1)-  N\,f^{\,2}(N)
=\{N+1\}- \{N\},
\label{eq:f-def2}
\end{eqnarray}
can be represented onto the Fock Hilbert space $F$ as follows
\begin{equation}
\fl A^{-}|0\rangle =0,\;\;  A^{-}|n\rangle=\sqrt{\{n\}}|n-1\rangle,
\;\;   A^{+}|n\rangle=\sqrt{\{n+1\}}|n+1\rangle,
\;\;   \{N\}|n\rangle=\{n\}|n\rangle,
\end{equation}
where the deformed number denoted by the symbol $\{n\}:=n\,f^{2}(n)$ is usually called the $f$-basic
number. One defines $\{0\}:=\lim_{n\to 0}n\,f^{2}(n)$
and the generalized factorial $\{n\}!:=\{n\}(\{n-1\}!)$
with $\{0\}!=1$  by convention.

Given the Hamiltonian (\ref{eq:fkeqm}), its energy spectrum can be worked out
by the usual tangent technique or by the quasideterminant approach \cite{qsd}.
The Hamiltonian Hilbert space ${\mathcal V}$
can be described by a direct sum of a finite set of  one dimensional complex spaces
${\mathcal V}_q$ generated by the orthonormalized states
$|E_{q}^{*}\rangle$, $q=0,1,\dots, k-1$, $k\geq 1$ and infinite dimensional complex space
 $\overline{\mathcal V}$ spanned by two towers of orthonormalized
states $|E^{\pm}_{n}\rangle$, $\;n\in \mathbb{N}$ such that $n+k\varepsilon\geq 0$.
Hence ${\mathcal V}=\oplus^{k-1}_{q=0}{\mathcal V}_q \oplus\overline{\mathcal V}$
and the reduced Hamiltonian (\ref{eq:fkeqm})
(omitting henceforth low indices) admits the spectral decomposition
\begin{eqnarray}
{\mathcal H}^{\rm red}=
\sum_{q=0}^{k-1}|E_q^{\ast}\rangle\,E_q^{\ast}
\,\langle E_q^{\ast}|\ +\
\sum_{n=0,\pm}^\infty\,
|E^\pm_{\widetilde{n}}\rangle\,E^\pm_{\widetilde{n}}\,
\langle E^\pm_{\widetilde{n}}| ,
\label{hamer}
\end{eqnarray}
where  $\widetilde{n}=n+n_0^{\varepsilon}$,  $n_0^{\varepsilon}:=\max(0,-k\,\varepsilon)$,
and, for $1 \leq q\leq k-1$,
\begin{eqnarray}
E_{q}^{*} = \frac{1}{2}
\left[(1+\epsilon-\varepsilon)\,\{q+1\}
+(1+\epsilon+\varepsilon\kappa)\,\{q\} \right],
\quad
|E_q^{\ast}\rangle =|q,-\varepsilon\rangle,
\label{eq:fground}
\end{eqnarray}
while the eigenenergies appearing in the infinite sum are defined as
\begin{eqnarray}
& {\mathcal E}(\{n\})= {\textstyle{\frac{1}{2}
\left[\frac{\epsilon}{2}\{n+ k\varepsilon+1\}
+\frac{1}{2}(1+\epsilon+\kappa)\{n+ k\varepsilon\} \right. }}\cr
&{\textstyle{ \left.  -\left(1+\frac{\epsilon}{2}\right)\{n+1\}
-\frac{1}{2}(1+\epsilon-\kappa)\{n\} \right]}} , \cr
&Q(\{n\})={\textstyle{ \left[{\mathcal E}^2(\{n\})\,+\,
|\lambda(n+k\varepsilon)|^{2}
\left(\frac{\{n+k\varepsilon\}!}{\{n\}!}\right)^{\varepsilon}\;
\right]^{\frac{1}{2}}}},
 \label{eq:EQ}\\
&\fl E^{\pm}_{n}={\textstyle{\frac{1}{2}\left[
\frac{\epsilon}{2}\{n+ k\varepsilon + 1\} +
\frac{1}{2}(1+\epsilon+\kappa)\{n+ k\varepsilon\}
+\left(1+\frac{\epsilon}{2}\right)\{n+1\} +
\frac{1}{2}(1+\epsilon-\kappa)\{n\}\right] }}\cr
&\fl \ \ \ \ \
\pm Q(\{n\}).
\label{eq:feigenval}
\end{eqnarray}
Given
\begin{eqnarray}
& \fl \sin\vartheta(\{n\})=\,e^{i\varphi_{\lambda}(n)}
\left[\frac{Q(\{n\})-{\mathcal
E}(\{n\})}{2Q(\{n\})}\right]^{\frac{1}{2}}, \quad
\cos\vartheta(\{n\})= \left[\frac{Q(\{n\})+{\mathcal
E}(\{n\})}{2Q(\{n\})} \; \right]^{\frac{1}{2}},
\label{eq:fmixang}
\end{eqnarray}
the eigenstates of the two towers can be expressed by
\begin{eqnarray}
&|E^{+}_{n}\rangle= \sin\vartheta(\{n\})\, |n,+\rangle +
\cos\vartheta(\{n\})\, |n+k\varepsilon,-\rangle,
\label{eq:feigenstatp}\\
&|E^{-}_{n}\rangle= \cos\vartheta(\{n\})\, |n,+\rangle -
\overline{\sin\vartheta(\{n\})}\, |n+k\varepsilon,-\rangle,
\label{eq:feigenstatm}
\end{eqnarray}
with $n\geq k$ for $\varepsilon =-$.
The notation $\overline{X}$ stands for the complex conjugate of the quantity $X$.
Note that $\Delta\, E_n= E_{n}^{+}-E_{n}^{-}$ $=2{Q}(\{n\})$
assigns the Zeeman spin splitting to the Rabi frequency up to a constant.

Allowing the passage from one basis of ${\mathcal V}$ to another, the operators
\begin{equation}
{\mathcal U}=\sum_{n=0,\pm}^\infty\,
|E^\pm_{\widetilde{n}}\rangle\,\langle n,\pm|,\qquad
{\mathcal U}^\dagger=\sum_{n=0,\pm}^\infty\,
|n,\pm\rangle\,\langle E^\pm_{\widetilde{n}}|
\label{eq:expuudag3}
\end{equation}
are mutually adjoint on  $\overline{\mathcal V}$ but non unitary on
${\mathcal V}$. We have the identities ${\mathcal U}^\dagger\;
{\mathcal U} =\mathbb{I}_{\mathcal V}$, ${\mathcal U}\; {\mathcal
U}^\dagger =
 \mathbb{I}_{\overline{\mathcal V}}$,
 ${\mathcal U}\,{\mathcal V}=\overline{\mathcal V}$,
${\mathcal U}^\dagger\,{\mathcal V}={\mathcal V}$ and
${\mathcal U}^\dagger\,\overline{\mathcal V}={\mathcal V}$.
Therefore, the reduced Hamiltonian (\ref{hamer}) can be diagonalized
in terms of
\begin{eqnarray}
\mathbb{H}^{\rm red}={\mathcal U}^\dagger\,{\mathcal H}^{\rm red}\,{\mathcal U}=
\sum_{n=0,\pm}^\infty\,|n,\pm\rangle\, E^\pm_{\widetilde{n}}\,\langle n,\pm|,
\label{eq:resspec}\\
{\mathcal U}\,\mathbb{H}^{\rm red}\,{\mathcal U}^\dagger=
\sum_{n=0,\pm}^\infty\,|E^\pm_{\widetilde{n}}\rangle\,
E^\pm_{\widetilde{n}}\, \langle E^\pm_{\widetilde{n}}|=
{\mathcal H}^{\rm red}\,-\,\sum_{q=0}^{k-1}
|E_q^{\ast}\rangle\,E_q^{\ast}\,
\langle E_q^{\ast}| .
\label{eq:reciresspec}
\end{eqnarray}
The following operators are adjoint of one another on ${\mathcal V}$,
\begin{eqnarray}
\fl \mathbb{M}^{-}=
\sum_{n=0,\pm}^\infty\,|n-1,\pm\rangle\, K_{\pm}(\{n\})\,\langle n,\pm|,
\quad
\mathbb{M}^{+}=
\sum_{n=0,\pm}^\infty\,|n+1,\pm\rangle\,
\overline{K_{\pm}(\{n+1\})}\,\langle n,\pm|,
\label{diagan}
\end{eqnarray}
with $K_{\pm}(\{n\})$ arbitrary complex functions
of $\{n\}$ such that $K_{\pm}(\{0\})=0$.
Reciprocally, the operators
${\mathcal M}^{-}=  {\mathcal U}\,\mathbb{M}^{-}\,{\mathcal U}^\dagger$
and ${\mathcal M}^{+}=  {\mathcal U}\,\mathbb{M}^{+}\,{\mathcal U}^\dagger$
are mutually adjoint on the subspace $\overline{{\mathcal V}}$
and have lowering and raising actions within each of the towers, namely
\begin{eqnarray}
&&
{\mathcal M}^{-}|E_q^{\ast}\rangle=0, \quad
{\mathcal M}^{+}|E_q^{\ast}\rangle=0,
\quad  q=0,1,\dots,k-1,\label{eq:qac}\\
&& {\mathcal M}^{-}|E^{\pm}_{\widetilde{n}}\rangle=
K_{\pm}(\{n\})\;|E^{\pm}_{\widetilde{n}-1}\rangle,
\qquad
{\mathcal M}^{+}|E^{\pm}_{\widetilde{n}}\rangle=
\overline{K_{\pm}(\{n+1\})}\;|E^{\pm}_{\widetilde{n}+1}\rangle.
\label{eq:nac}
\end{eqnarray}

The generalized annihilation operator eigenvalue problem
\begin{eqnarray}
{\mathcal M}^-|z;\tau_\pm;\theta,\phi\rangle =
 z\,\widetilde{\mathbb{Q}}_{\mathcal V}\,|z;\tau_\pm;\theta,\phi\rangle,
 \label{eq:fevpb}\\
\widetilde{\mathbb{Q}}_{\cal V}:=
\sum_{q=0}^{k-1}|E_q^{\ast}\rangle\,\langle E_q^{\ast}|\ +\
\sum_{n=0,\pm}^\infty\,|E^{\pm}_{\widetilde{n}}\rangle\,
h_{f}^{\pm}(n)\,\langle E^{\pm}_{\widetilde{n}}|,
\end{eqnarray}
where the quantities $h_{f}^{\pm}(n)\ne 0$ are such that
$h_{f}^{\pm}(n) \to 1$ as $f(N)\to 1$, should be solved in order
to define the $(k,\varepsilon,\kappa,f)$-VCS denoted by
$|z;\tau_\pm;\theta,\phi\rangle$. The terminology {\it generalized} is justified by
the fact that the eigenvalue problem is stated here through  generalized deformed
arbitrary structure
functions $K_{\pm}$ and $h_f^{\pm}$, incorporated in ${\mathcal M}^-$
 and $\widetilde{\mathbb{Q}}_{\mathcal V}$, respectively,
 and encompasses known
related eigenvalue problems \cite{hus} (the matrix eigenvalue problem developed
 in the sequel includes the models in \cite{al} (see also references therein)).
As the most simple instance,  at the canonical limit
$\lim_{f(N)\to 1} \widetilde{\mathbb{Q}}_{\cal V}\equiv \mathbb{I}_{\cal V}$ and
the problem (\ref{eq:fevpb}) exactly
reduces to a Barut-Girardello eigenvalue problem for VCSs.
In addition, the parameters $\tau_\pm$
are introduced for the Gazeau-Klauder axiom of temporal stability, $(\theta,\phi)$
parametrize the unit vectors of the sphere  $S^2$ and also
determine the vector feature of the NVCSs $|z;\tau_\pm;\theta,\phi\rangle$.

The general form of the
$(k,\varepsilon,\kappa,f)$-VCSs fulfilling all axioms of Gazeau-Klauder
are given by \cite{BGH2}
\begin{eqnarray}
\fl |z;\tau_\pm;\theta,\phi\rangle =
\ \ \ {\cal N}^+(|z|)\,\cos\theta\,
\sum_{n=0}^\infty\, \frac{z^n}{K^0_+(\{n\})!}\,
(h_{f}^{+}(n-1)!)h_{f}^{+}(0)\,
e^{-i\omega_0\tau_+\,E^+_{\widetilde{n}}}\,|E^+_{\widetilde{n}}\rangle  \cr
\fl + \ {\cal N}^-(|z|)\,e^{i\phi}\,\sin\theta\,
\sum_{n=0}^\infty \,
\frac{z^{n}}{K^0_-(\{n\})!}\,(h_{f}^{-}(n-1)!)h_{f}^{-}(0) \,
e^{-i\omega_0\tau_-\,E^-_{\widetilde{n}}}\,|E^-_{\widetilde{n}}\rangle,
\label{eq:fcoh2}\\
{\cal N}^\pm(|z|)= \left[ \sum_{n=0}^\infty\,
\frac{|z|^{2n}}{\left(K^0_\pm(\{n\})!\right)^2}
\left((h_{f}^{\pm}(n-1)!)h_{f}^{\pm}(0)\right)^{2} \right]^{-1/2}
\label{eq:fnorm}\\
\end{eqnarray}
where ${\cal N}^\pm(|z|)$ are normalization factors, the real
positive functions $K^0_\pm(\{n\}):=|K_\pm(\{n\})|$ and
$h_{f}^{\pm}(n)$ are yet to be specified;
$K^0_\pm(\{n\})!:=\prod_{k=1}^{n}K^0_\pm(\{k\})$,
$h_{f}^{\pm}(n-1)!:=\prod_{k=1}^{n-1} h_{f}^{\pm}(k)$, with by
convention $K^0_\pm(\{0\})!=1$, $h_{f}^{\pm}(0)!=1$ and
$h_{f}^{\pm}(-1)!=(h_{f}^{\pm}(0))^{-1}$. The convergence radii of
${\cal N}^\pm(|z|)$ are $R_\pm=\lim_{n\to\infty}(K^0_\pm(\{n\})/
h_{f}^{\pm}(n-1))$ and depend on the choice of $K^0_\pm(\{n\})$ as
well as on $h_{f}^{\pm}(n)$. Consequently, the NVCSs of this form
live in the disc $|z|\leq R$, $R=\min\left(R_+,R_-\right)$.

The heretofore developments meet also a matrix formulation.
Given the $2\times 2$ diagonal matrix
$K(\{n\}):= {\rm diag}(K_+(\{n\}),\;K_-(\{n\}) )$
and $\overline{K(\{n\})}$ its adjoint, the diagonal matrix annihilation
operator connected with (\ref{diagan}) is
\begin{eqnarray}
\mathbb{M}^{-}= \sum_{n=0,\pm}^\infty\, K(\{n\})
|n-1,\pm\rangle\,\langle n,\pm|,
\end{eqnarray}
and again ${\mathcal M}^{-}={\cal U}\,\mathbb{M}^{-}\,{\cal U}^\dagger$.
We translate the eigenvalue problem (\ref{eq:fevpb}) equally as
\begin{eqnarray}
{\mathcal M}^-|(z,w);\tau_\pm;\pm\rangle =
\widetilde{\mathfrak{Z}}(z,w)\,\widetilde{\mathbb{Q}}_{\cal V}\,
|(z,w);\tau_\pm;\pm\rangle,
\label{eq:ev1}
\end{eqnarray}
where $\widetilde{\mathfrak{Z}}(z,w)$ is a $2\times 2$ matrix operator
of the two complex variables $(z,w)$, $\pm$ is the
spin-vector dependence  now replacing the $S^2$ unit sphere vectors.
Assuming that $\mathfrak{Z}={\cal U}^\dagger\;\widetilde{\mathfrak{Z}} \;{\cal U}$
is a complex constant matrix and defining
$\mathbb{Q}_{\cal V}:={\cal U}^\dagger \widetilde{\mathbb{Q}}_{\cal V} \;{\cal U}$,
we rephrase (\ref{eq:ev1}) into the diagonal basis as
\begin{eqnarray}
\mathbb{M}^-|\mathfrak{Z};\tau_\pm;\pm\rangle =
\mathfrak{Z}\,\mathbb{Q}_{\cal V}\,
|\mathfrak{Z};\tau_\pm;\pm\rangle,\quad
|\mathfrak{Z};\tau_\pm;\pm\rangle = {\cal U}^\dagger\;
|(z,w);\tau_\pm;\pm\rangle,\label{eq:ev2} \\
\mathbb{Q}_{\cal V}=
\sum_{n=0,\pm}^\infty\,  h_f(n) |n,\pm\rangle\,\langle n,\pm|,
\quad h_f(n)={\rm diag}(h_f^+(n),h_f^-(n)) .
\nonumber
\end{eqnarray}
The problem (\ref{eq:ev2}) is a matrix eigenvalue problem.
A canonical class of these problems can be exactly carried out
if ${\mathfrak{Z}}$ is in a normal
or in a quaternionic matrix domain \cite{al}.
In the particular instance of a nonlinear model, assuming also
that ${\mathfrak{Z}}={\rm diag}(z,w)$,
the general Gazeau-Klauder NVCSs solution of the eigenvalue problem
(\ref{eq:ev2}) is given by
\begin{eqnarray}
|\mathfrak{Z};\tau_\pm;\pm\rangle
=N(\mathfrak{Z})\sum_{n=0}^{\infty}\,
R_0(n)\exp[-i\omega_0\tau E_{\widetilde{n}}]\;
\mathfrak{Z}^{n}\,|n,\pm\rangle,
\label{eq:vcs1}\\
N(\mathfrak{Z})^{-2}=\sum_{n=0}^{\infty}\,
\left(|z|^{2n}(R^0_{+}(n))^2 + |w|^{2n}(R^0_-(n))^2\right),
\label{eq:norm}\\
R^0(n)={\rm diag}(R^0_+(n) ,R^0_-(n))
=(K^0(\{n\})!)^{-1}(h_f(n-1)!h_f(0))
\end{eqnarray}
where $N(\mathfrak{Z})$ is the normalization factor,
$\tau={\rm diag}(\tau_+,\tau_-)$ and
$E_{\widetilde{n}}={\rm diag}(E^+_{\widetilde{n}},E^-_{\widetilde{n}})$.
Note that the convergence radii of the series (\ref{eq:norm})
are such that $|z|\leq L_+$, $|w|\leq L_-$ and
$L_\pm=\lim_{n\to\infty}K^0_\pm(\{n\})/h^\pm_f(n-1)$.

Given in the form (\ref{eq:fcoh2}) or (\ref{eq:vcs1}), the NVCSs
contain still some undetermined quantities, hence the name of
general form of NVCSs. They have been built with respect to three
over four Gazeau-Klauder axioms: the continuity in labeling, the
temporal stability, the normalizability. It then remains the
resolution of the identity. It turns out that explicit examples of
deformation of NVCSs in both forms have exact solutions to their
Stieljes moment problem by further constraining the operator algebra
as $[{\mathcal M}^-,{\mathcal M}^+]= \mathbb{I}_{\overline{\cal
V}}$, or by the action identity constraint \cite{kl2} in the case of
the canonical limit only \cite{BGH1,BGH2}. Hence, generalized
solvable NVCSs meeting all requirements of Gazeau-Klauder have been
successfully built from a deformed physical model.

\section{Hamiltonian Hilbert space organization and ladder operators}
\label{sect3}

From this section, we start our main new results dealing
with an extension of the previous analysis
of $(k,\varepsilon, \kappa, f)$-VCSs by a prolongation of
the annihilation operator ${\mathcal M}^-$ onto the finite dimensional
space generated by $\oplus_{q=0}^{k-1}{\mathcal V}_q$.
A number  of deformed displacements
for $S^2$, matrix NVCSs and issues about deformed $T$ operators will be
addressed in the next sections. The results of the remaining sections are totally dependent of the
Hilbert space organization that we set in this section.
We mention that, albeit quantities and operators may differ,
same notations as in Section \ref{sect2} will be used hereafter.

Let us first redefine the eigenstates and eigenenergies as
\begin{eqnarray}
|e_n^{-}\rangle = |E_n^{\ast}\rangle, \quad e_n^{-} =E_n^{\ast} ,\quad   n=0,1,\ldots,k-1,\\
|e_{n}^{-}\rangle = |E^{-}_{n-n^{-\varepsilon}_0}\rangle, \quad
e_{n}^{-} =E^{-}_{n-n^{-\varepsilon}_0},
 \quad n=k,k+1,\ldots\\
|e_{n}^{+}\rangle = |E^{+}_{n+n^{\varepsilon}_0}\rangle, \quad
e_{n}^{+}=E^{+}_{n+n^{\varepsilon}_0},  \quad n=0,1,\ldots,
 \label{redef}
 \end{eqnarray}
so that we obtain two new towers of states $|e_{n}^{\pm}\rangle$,
$n=0,1,\ldots$. Note that, the finite sequence $|E_n^{\ast}\rangle$, $n\leq k-1$,
could also be added to the tower $|e_n^{+}\rangle$ without loss of generality.

The spectral decomposition of the Hamiltonian (\ref{hamer}) appears
in the simpler form
\begin{eqnarray}
{\mathcal H}^{\rm red}=
\sum_{n=0,\pm}^\infty\,
|e^\pm_{n}\rangle\,e^\pm_{n}\,
\langle e^\pm_{n}|.
\label{hamer2}
\end{eqnarray}
Next the passage operators ${\mathcal U}$ and ${\mathcal U}^\dag$,
such that ${\cal U}|n,\pm\rangle=|e^\pm_{n}\rangle$ and
${\cal U}^\dagger|e^\pm_{n}\rangle=|n,\pm\rangle$, encounter the expansion
\begin{equation}
{\cal U}=\sum_{n=0,\pm}^\infty\,
|e^\pm_{n}\rangle\,\langle n,\pm|,\qquad
{\cal U}^\dagger=\sum_{n=0,\pm}^\infty\,
|n,\pm\rangle\,\langle e^\pm_{n}| .
\label{eq:expuudag4}
\end{equation}
They come now as unitary operators on ${\mathcal V}$ since they satisfy
\begin{eqnarray}
{\mathcal U}\,{\mathcal V}={\mathcal V},\quad
{\cal U}^\dagger\,{\cal V}={\cal V},\quad
{\cal U}^\dagger\; {\cal U} =  \mathbb{I}_{\cal V}= {\cal U}\; {\cal U}^\dagger.
\label{eq:idenuudag}
\end{eqnarray}
Therefore the Hamiltonian ${\cal H}^{\rm red}$ can be written
in a diagonal form
\begin{eqnarray}
\mathbb{H}^{\rm red} ={\mathcal U}^\dagger{\mathcal H}^{\rm red}{\cal U} =
\sum_{n=0,\pm}^\infty\,
|n,\pm\rangle\,e^\pm_{n}\, \langle n,\pm|.
\label{hamerdiag}
\end{eqnarray}
Conversely, in contrast with (\ref{eq:reciresspec}), we have ${\cal
U} \mathbb{H}^{\rm red}{\cal U}^\dagger ={\mathcal H}^{\rm red}$.
Thus this notation improves the understanding of the Hamiltonian
Hilbert space with a basis $|e^\pm_n\rangle$ mapped  onto the
diagonal basis $|n,\pm\rangle$ via a unitary operator ${\mathcal
U}^\dag$.

Now, let us focus on the annihilation and creation operators.
To begin with, consider
the diagonal and mutually adjoint operators
\begin{equation}
\fl \mathbb{M}^{-}=\sum_{n=0,\pm}^{\infty}\,| n-1,\pm\rangle\;
K_{\pm}(\{n\})\;\langle n,\pm|,
\;\;
\mathbb{M}^{+}= \sum_{n=0,\pm}^{\infty}\,|n+1,\pm\rangle\;
\overline{K_{\pm}(\{n+1\})}\;\langle n,\pm|,
\label{diagan2}
\end{equation}
where again $K_{\pm}(\{n\})$ are some functions of the $f$-basic
number $\{n\}$ with initial value $K_{\pm}(\{0\})=0$. Changing the
basis, the corresponding operators
\begin{eqnarray}
&& {\cal U} \mathbb{M}^{-}{\cal U}^\dagger = {\mathcal M}^{-}
=\sum_{n=0,\pm}^{\infty}\,| e^{\pm}_{n-1}\rangle\;
K_{\pm}(\{n\})\;\langle e^{\pm}_{n}|,
\label{eq:ann}\\
&&{\cal U} \mathbb{M}^{+}{\cal U}^\dagger = {\mathcal M}^{+} = \sum_{n=0,\pm}^{\infty}\,|e^{\pm}_{n+1}\rangle\;
\overline{K_{\pm}(\{n+1\})}\;\langle e^{\pm}_{n}|,
\label{eq:crea}
\end{eqnarray}
with lowering and raising actions as, $\forall n\in \mathbb{N}$,
\begin{eqnarray}
{\mathcal M}^{-}|e^{\pm}_{n}\rangle=
K_{\pm}(\{n\})\;|e^{\pm}_{n-1}\rangle,
\qquad
{\mathcal M}^{+}|e^{\pm}_{n}\rangle=
\overline{K_{\pm}(\{n+1\})}\;|e^{\pm}_{n+1}\rangle,
\end{eqnarray}
define mutually adjoint ladder operators on ${\cal V}$.

In terms of the ``old" basis
$\{|E_q^{\ast}\rangle, q=0,1,\ldots,k-1\}\cup \{|E^{\pm}_{\tilde{n}}\rangle, n=0,1,\ldots\}$, we have
\begin{eqnarray}
&&\fl {\mathcal M}^{-}|E_0^{\ast}\rangle =0,\qquad
{\mathcal M}^{+}|E_0^{\ast}\rangle =K_{-}(\{1\})|E_{1}^{\ast}\rangle,\\
&&\fl
{\mathcal M}^{-}|E_q^{\ast}\rangle=K_{-}(\{q\})|E_{q-1}^{\ast}\rangle, \quad
{\mathcal M}^{+}|E_q^{\ast}\rangle=K_{-}(\{q+1\})|E_{q+1}^{\ast}\rangle,
\quad  q=1,\dots,k-2,\\
&&\fl {\mathcal M}^{-}|E_{k-1}^{\ast}\rangle=K_{-}(\{k-1\})|E_{k-2}^{\ast}\rangle,
 \qquad
{\mathcal M}^{+}|E_{k-1}^{\ast}\rangle=K_{-}(\{k\})|E_{\widetilde{0}}^{-}\rangle,\\
&& \fl {\mathcal M}^{-}|E_{\widetilde{0}}^{-}\rangle=K_{-}(\{k\})|E_{k-1}^{\ast}\rangle,\qquad
 {\mathcal M}^{+}|E_{\widetilde{0}}^{-}\rangle=K_{-}(\{1\})|E_{\widetilde{1}}^{-}\rangle \\
&& \fl {\mathcal M}^{-}|E^{\pm}_{\widetilde{n}}\rangle=
K_{\pm}(\{n\})\;|E^{\pm}_{\widetilde{n}-1}\rangle,
\qquad
{\mathcal M}^{+}|E^{\pm}_{\widetilde{n}}\rangle=
\overline{K_{\pm}(\{n+1\})}\;|E^{\pm}_{\widetilde{n}+1}\rangle, \;\;n\geq 1,
\end{eqnarray}
which are to be compared with (\ref{eq:qac}) and (\ref{eq:nac}). The
action of the annihilation operator is therefore ensured between the
two parts of the Hilbert space $\oplus_{q=0}^{k-1}{\mathcal V}_q$
and $\overline{\mathcal  V}$ by the operator $
|E_{k-1}^{\ast}\rangle K_{-}(\{k\}) \langle E_{\widetilde{0}}^{-}| $
(and its adjoint in the case of raising action). It could be also
defined over type of prolongation of annihilation operator using the
mapping $| E_{0}^{\ast} \rangle \,\langle E_{k-1}^{\ast}|$ and
entailing a kind of ``cyclic" annihilation operator onto
$\oplus_{q=0}^{k-1}{\mathcal V}_q$. This type of annihilation
operator proves to be well defined for finite dimensional Hilbert
spaces \cite{kuang,leons}. However, restricted to that latter
situation,
 some difficulties emerge in the construction of CSs as we shall define it later.

Finally, we give the matrix entries of the annihilation operator ${\mathcal M}^-$ (\ref{eq:ann})
in the diagonal basis $\{|n,\pm\rangle, \,\, n\in \mathbb{N} \}$
\begin{eqnarray}
&& \fl {\cal M}^{-}=
\sum_{q=0}^{k-1}\,
|q-1,-\varepsilon \rangle\,{\mathcal M}^-_{-\varepsilon \,-\varepsilon}(\{q\})\,\langle q,-\varepsilon | \cr\cr
&&\fl  +
|k-1,-\varepsilon\rangle\,{\cal M}^-_{-\varepsilon\, +}(\{k\})\,\langle \widetilde{0},+| +
|k-1,-\varepsilon\rangle\,{\cal M}^-_{-\varepsilon\, -}(\{k\})\,
\langle \widetilde{0}+k\varepsilon,-| \cr &&\\
&&\fl +\sum_{n=0}^\infty\,
|\widetilde{n},+\rangle\,{\cal M}^-_{++}(\{n\})\,\langle \widetilde{n}+1,+| \ + \
\sum_{n=0}^\infty\,|\widetilde{n},+\rangle\,{\cal M}^-_{+-}(\{n\})\,\langle \widetilde{n}+k\varepsilon+1,-| \cr
 &&\fl +
\sum_{n=0}^\infty\,|\widetilde{n}+k\varepsilon-1,-\rangle\,{\cal M}^-_{-+}(\{n\})\,
\langle \widetilde{n},+| \ + \
\sum_{n=0}^\infty\,|\widetilde{n}+k\varepsilon-1 ,-\rangle\,{\cal M}^-_{--}(\{n\})\,
\langle \widetilde{n}+k\varepsilon,-|
\nonumber
\end{eqnarray}
where
\begin{eqnarray}
&&\fl {\mathcal M}^-_{-\varepsilon \,-\varepsilon}(\{0\})=0,\;\:
{\mathcal M}^-_{-\varepsilon \,-\varepsilon}(\{q\}) = K_{-}(\{q\}),\quad
q=1,\ldots,k-1,\cr
&& \fl {\cal M}^-_{-\varepsilon\, +}(\{k\})=
K_{-}(\{k\}) \cos\vartheta(\{\widetilde{0}\}), \quad
{\cal M}^-_{-\varepsilon\, -}(\{k\}) =
- K_{-}(\{k\}) \sin\vartheta(\{\widetilde{0}\}),
\cr\cr
&&\fl {\cal M}^-_{++}(\{n\})=
\sin\vartheta(\{ \widetilde{n}\})
\overline{\sin\vartheta(\{\widetilde{n}+1\})}K_{+}(\{n+1\})
\cr &&
+ \cos\vartheta(\{\widetilde{n}\})\cos\vartheta(\{\widetilde{n}+1\})K_{-}(\{n+1\}),
\cr \cr
&&\fl {\cal M}^-_{+-}(\{n\})=
\sin\vartheta(\{\widetilde{n}\})\cos\vartheta(\{\widetilde{n}+1\})\,K_{+}(\{n+1\})
\cr
&& -
\cos\vartheta(\{\widetilde{n}\})\sin\vartheta(\{\widetilde{n}+1\})\,K_{-}(\{n+1\}),\cr
\cr
&&\fl  {\cal M}^-_{-+}(\{n\})=
\cos\vartheta(\{\widetilde{n}-1\})\,\overline{\sin\vartheta(\{\widetilde{n}\})}K_{+}(\{n\})-
\overline{\sin\vartheta(\{\widetilde{n}-1\})}\cos\vartheta(\{\widetilde{n}\})K_{-}(\{n\}),\cr
&& \cr
&&\fl
{\cal M}^-_{--}(\{n\})=
\cos\vartheta(\{\widetilde{n}-1\})\cos\vartheta(\{\widetilde{n}\})\,K_{+}(\{n\})+
\overline{\sin\vartheta(\{\widetilde{n}-1\})}\sin\vartheta(\{\widetilde{n}\})K_{-}(\{n\}),
\label{eq:andiag}\cr&&
\end{eqnarray}
with ${\cal M}^-_{-+}(\{0\})=0={\cal M}^-_{--}(\{0\})$, for $K_\pm(0)=0$.
Taking the complex conjugate of these expressions leads to the analogue
of these quantities connected with the raising operator.
The parameter functions $K_\pm(\{n\})$ again indicate the freedom in the choice of
creation and annihilation operators that we will restrict in order to allow the existence
of NVCSs according to a series of general axioms \cite{gk}.
As was observed in \cite{BGH2}, the current Hilbert space structure
and its organization do not depend on the quantization scheme.
Thus the above results remain valid for the undeformed canonical situation
recovered by setting $\{n\} \to n$ in all expressions.

\section{Nonlinear vector coherent states}
\label{sect4}

In this section, we construct  classes of Gazeau-Klauder NVCSs using the  annihilation operator eigenvalue
 problem of the forms
(\ref{eq:fevpb}) and (\ref{eq:ev1}) resolved now within the new Hilbert space structure.
The Hamiltonian expectation value and the Rabi oscillations for these states are computed and
 exact solutions to their overcompleteness problem are given. The corresponding matrix formulation,
deformed displacement operators,
 dual NVCSs and $T$ operators will be consistently defined in the next sections.

\subsection{Identifying  NVCSs}
\label{subsect41}

We proceed to the construction of NVCSs with respect to the set of Gazeau-Klauder axioms
by solving the generalized eigenvalue problem
\begin{eqnarray}
{\mathcal M}^-|z;\tau_\pm;\theta,\phi\rangle =
 z\,\widetilde{\mathbb{Q}}_{\cal V}\,|z;\tau_\pm;\theta,\phi\rangle,
 \label{eq:fevpb2}\\
\widetilde{\mathbb{Q}}_{\cal V}:=
\sum_{n=0,\pm}^\infty\,|e^{\pm}_{n}\rangle\,
h_{f}^{\pm}(n)\,\langle e^{\pm}_{n}|,
\end{eqnarray}
where the parameters $\tau_\pm$,
$\theta$ and $\phi$ are introduced below and the functions $h_f^{\pm}(n)$
have the same properties as stated before.
Assuming that the solution of the problem (\ref{eq:fevpb2}) is of the
form
\begin{eqnarray}
|z;\tau_\pm;\theta,\phi\rangle=
\sum_{n=0,\pm}^{\infty} C^{\pm}_{n}(z)|e^{\pm}_{n}\rangle,
\label{eq:series}
\end{eqnarray}
where $C^{\pm}_{n}(z)$ are complex continuous functions of the complex
variable $z$, then substituting (\ref{eq:series}) into (\ref{eq:fevpb2}) leads to
the recurrence relation
\begin{equation}
\forall n\in \mathbb{N},\quad
C^\pm_{n+1}(z)K_\pm(\{n+1\}) = z\,h_{f}^{\pm}(n)\,C^\pm_n(z).
\label{eq:reccurel}
\end{equation}
A simple solution of this recurrence is
\begin{equation}
C_{n}^{\pm}(z)=\frac{z^n}{K_\pm(\{n\})!}(h_{f}^{\pm}(n-1)!)
h_{f}^{\pm}(0)\,C^\pm_0(z),\qquad n\geq 1,
\label{eq:fresoreccur}
\end{equation}
with $C^\pm_0(z)$ arbitrary continuous
complex functions of $z$. The sense of generalized factorials
remains as $K_{\pm}(\{n\})!:= \prod_{p=1}^{n}K_{\pm}(\{p\})$,
$h_{f}^{\pm}(n-1)!=\prod_{p=1}^{n-1} h_{f}^{\pm}(p)$ with,
by convention, $K_{\pm}(\{0\})!=1$, $h_{f}^{\pm}(0)!=1$ and
$h_{f}^{\pm}(-1)!=(h_{f}^{\pm}(0))^{-1}$.
Therefore, the expression of $C_{n}^{\pm}(z)$
(\ref{eq:fresoreccur}) is still correct for $n\geq 0$.

{\it Temporal stability} condition refers to the CS stability under
time evolution operator $U(t)=\exp(-i\omega_0\,t\,{\cal H}^{\rm red})$.
In other words, the CSs should transform into one another under time translations.
More subtle considerations about this notion can be found in \cite{rok3}.
 The latter axiom can be reached by introducing a phase
$\varphi_\pm(\{n\})$ such as $K_\pm(\{n\}) = \exp[i\,\varphi_\pm(\{n\})] K^0_\pm(\{n\})$,
 $K^0_\pm(\{n\})$ being real positive quantities, and after
imposing the relations, for all $n=1,2,\cdots$,
\begin{equation}
\varphi_\pm(\{n\})=
\omega_0\tau_\pm\left[e^{\pm}_{n} -e^{\pm}_{n-1}\right],
\end{equation}
where $\tau_\pm$ is a new parameter. One has to define
$C^\pm_0(z)= {\cal C}^{\pm}_0(z)
\exp[-i\,\omega_0\,\tau_{\pm}\, e^{\pm}_{0}]$ so that
the problem (\ref{eq:fevpb2}) gives the solution
\begin{equation}
\fl |z;\tau_\pm;\theta,\phi\rangle = \sum_{n=0,\pm}^\infty\,
\frac{z^n}{K^0_\pm(\{n\})!}\,(h_{f}^{\pm}(n-1)!)\,h_{f}^{\pm}(0)\;
{\cal C}^\pm_0(z)\,
e^{-i\omega_0\tau_\pm\,e^\pm_{n}}\,|e^\pm_{n}\rangle
\label{eq:fcoh}
\end{equation}
with the property
\begin{equation}
U(t)|z;\tau_\pm;\theta,\phi\rangle = |z;\tau_\pm+t;\theta,\phi\rangle.
\label{eq:time}
\end{equation}
At this stage, we mention that the construction adopted here to
ensure the temporal stability requirement is equivalent to the
procedure by Roknizadeh and Tavassoly \cite{rok3} where an evolution
operator maps any generalized CS to a temporal stable one.
 We also point out the fact that the anterior problems of a
singular state associated to the eigenvalue $z=0$
\cite{hus,BGH1,BGH2} has been totally removed. For instance, in the
case $k=1$, i.e. the so called Jaynes-Cummings model, the states
$|E^*\rangle$ and $|E^\pm_0\rangle$
  share same eigenvalue $z=0$. However,
the state $|E^*\rangle$ is not included in the definition of the
CSs; then a singularity breaks the Gazeau-Klauder axiom of {\it
continuity of labeling} in $z$ for the eigenvalue problem
(\ref{eq:fevpb}). In the case of the $k$-multiphoton model, $k$
states violate the latter axiom. This difficulty is generally
circumvented by the simple claim that the vectors with eigenvalue
$z=0$ may be expressed as combination of the eigenstates
$|E^+_{n_0^\varepsilon}\rangle$ and $|E^-_{n_0^\varepsilon}\rangle$.
Here such an issue is totally avoided due to the ``continuous"
action of the annihilation operator, passing from the infinite towers to the
finite dimensional part of the Hamiltonian Hilbert space.

Defining
\begin{equation}
{\cal C}^+_0(z)={\cal N}^+(|z|)\,\cos\theta,\qquad
{\cal C}^-_0(z)={\cal N}^-(|z|)\,e^{i\phi}\,\sin\theta,
\end{equation}
and introducing the $S^2$ unit vector coordinates $(\theta,\phi)$
and the functions
\begin{equation}
{\cal N}^\pm(|z|)=
\left[
\sum_{n=0}^\infty\,
\frac{|z|^{2n}}{\left(K^0_\pm(\{n\})!\right)^2}
\left((h_{f}^{\pm}(n-1)!)h_{f}^{\pm}(0)\right)^{2}
\right]^{-1/2}
\label{eq:fnorm2}
\end{equation}
of convergence radii $R_\pm$
\begin{equation}
R_\pm=\lim_{n\to\infty}\left[
\frac{K^0_\pm(\{n\})}{h_{f}^{\pm}(n-1)}\right],
\label{eq:fradii}
\end{equation}
ensure the {\it normalization axiom}.
Finally, the general $(k,\varepsilon,\kappa,f)$-VCS associated
with the  generalized spin-orbit model (\ref{hamer2}) has the form
\begin{eqnarray}
&&\fl |z;\tau_\pm;\theta,\phi\rangle =
\ \ \ {\cal N}^+(|z|)\,\cos\theta\,
\sum_{n=0}^\infty\, \frac{z^n}{K^0_+(\{n\})!}\,
(h_{f}^{+}(n-1)!)h_{f}^{+}(0)\,
e^{-i\omega_0\tau_+\,e^+_{n}}\,|e^+_{n}\rangle  \cr
&&\fl \ \ \ \ \ + \ {\cal N}^-(|z|)\,e^{i\phi}\,\sin\theta\,
\sum_{n=0}^\infty \,
\frac{z^{n}}{K^0_-(\{n\})!}\,(h_{f}^{-}(n-1)!)h_{f}^{-}(0) \,
e^{-i\omega_0\tau_-\,e^-_{n}}\,|e^-_{n}\rangle,
\label{eq:fcoh21}
\end{eqnarray}
with $|z|\leq R$, $R=\min\left(R_+,R_-\right)$; the positive
functions $K^0_\pm(\{n\})$ and $h_{f}^{\pm}(n)$ are to be specified.
One notices the similar structure of NVCSs (\ref{eq:fcoh21}) and
(\ref{eq:fcoh2}). However, these two classes of CSs radically differ
since they have not built onto the same Hilbert space.

The {\it overcompleteness condition} is a necessary axiom
that any family of CSs ought to satisfy \cite{al,gk,kl2}.  For the
NVCSs (\ref{eq:fcoh21}), this condition can be formulated as
\begin{equation}
\mathbb{I}_{\cal V}=
\sum_{n=0,\pm}^\infty\,|e^\pm_{n}\rangle\,\langle e^\pm_{n}|=
\int_{D_R\times S^2}\,d\mu(z;\theta,\phi)\,
|z;\tau_\pm;\theta,\phi\rangle\,\langle z;\tau_\pm;\theta,\phi|,
\label{eq:resol}
\end{equation}
with the $SU(2)$ matrix-valued integration measure over $D_R\times S^2$, $d\mu(z;\theta,\phi)$ admitting the
parametrization as
\begin{equation}
\fl d\mu(z;\theta,\phi)=d^2z\,d\theta\,\sin\theta\,d\phi\,
\left\{{\cal W}^+(|z|)\sum_{n=0}^\infty|e^+_{n}\rangle\langle e^+_{n}|\,+\,
{\cal W}^-(|z|)\sum_{n=0}^\infty|e^-_{n}\rangle\langle e^-_{n}|\right\},
\label{eq:weight}
\end{equation}
where ${\cal W}^\pm(|z|)$ are yet unknown real weight functions. A
direct substitution in (\ref{eq:resol}), using the radial
parametrization $z=r\,\exp(i\varphi)$, so that $d^2z=r\,dr
\,d\varphi$, where $r\in[0,R)$ and $\varphi\in[0,2\pi[$, leads to
the Stieljes moment problems
\begin{equation}
\forall \,n\in\mathbb{N},\qquad\;\;\int_0^{R^2} du\,u^n\,h^\pm(u)= \,\frac{\left(K^0_\pm(\{n\})!\right)^2}
{\left((h_{f}^{\pm}(n-1)!)h_{f}^{\pm}(0)\right)^2},
\label{eq:stj4}
\end{equation}
where $u=r^2$ and the functions $h^\pm(r^2)$ are
\begin{equation}
h^+(r^2)=\frac{4\pi^2}{3}\,|{\mathcal N}^+(r)|^2\,{\cal W}^+(r),\qquad
h^-(r^2)=\frac{8\pi^2}{3}\,|{\mathcal N}^-(r)|^2\,{\cal W}^-(r).
\label{eq:hfunctions}
\end{equation}
Let us remark that the problems (\ref{eq:stj4}) have the same form
of the Stieljes problems as was developed in \cite{BGH2}.
Consequently, the same techniques as found therein could be used
in order to determine a solvable and deformed resolution of the identity.
Some explicit solutions will be furnished later.
Provided these answers, we may assume that,
there exists a wide class of sets of $(k,\varepsilon,\kappa,f)$-VCSs fulfilling
the Gazeau-Klauder axioms in a nonempty disk $D_R$ of $\mathbb{C}$
distinguishable by specific choices of $K^0_\pm(\{n\})>0$ and $h_{f}^\pm(n)$,
for instance.

\subsection{Some expectation values and action-angle variables}
\label{subsect42}
We briefly give the expectation value of the Hamiltonian
operator and discuss the atomic spin time evolution average.

The Hamiltonian mean value measured in any state is
given by
\begin{eqnarray}
&&\fl \langle{\cal H}^{\rm red}\rangle
=\ \ \ |{\mathcal N}^+(|z|)|^2\,\cos^2\theta\,
\sum_{n=0}^\infty\,
\frac{|z|^{2n}}{\left(K^0_+(\{n\})!\right)^2}\,
\left((h_{f}^{+}(n-1)!)h_{f}^{+}(0)\right)^{2}\,
e^+_{n}\cr
&&\fl  +\ |{\mathcal N}^-(|z|)|^2\,\sin^2\theta\,
\sum_{n=0}^\infty\,
\frac{|z|^{2n}}{\left(K^0_-(\{n\})!\right)^2}\,
\left((h_{f}^{-}(n-1)!)h_{f}^{-}(0)\right)^{2}\,
e^-_{n}.
\label{eq:fhexpct}
\end{eqnarray}
The average spin time evolution (atomic inversion in quantum optics)
is defined by $\langle\sigma_3(t)\rangle=\langle U^{-1}(t)\sigma_3 U(t)\rangle$,
with the time evolution operator
$U(t)=\exp(-i\omega_0 t\,{\mathcal H}^{\rm red})$.
We get a similar atomic inversion as the one of \cite{BGH2}
with Rabi oscillation currently of the form
\begin{eqnarray}
\Psi_{n}(t)&=&
\omega_0\left[\left(t+\tau_+\right)e^+_{n}\,-\,
\left(t+\tau_-\right)e^-_{n}\right]\,
+\,\phi-\varphi_{\lambda}(n) \cr
&=&
\omega_0\Delta e_{n}\,t\,
+\,\omega_0\left[\tau_+ e^+_{n} - \tau_- e^-_n\right]
+\phi-\varphi_{\lambda}(n)
\end{eqnarray}
showing an explicit time dependence due to the mixed-spin matrix
elements when $\lambda(\{N\})\ne 0$ (in the limit $\lambda(\{N\})\to
0$, oscillations collapse) and consisting only of Rabi oscillations
as expected. This is  a general property for quantum optics system
prepared in a CS of the radiation field \cite{hus4}.

Klauder's argument \cite{kl2} that the stability of CSs under time evolution
could be translated by defining continuous parameters, so-called canonical action-angle variables $(J, \tau)$, should be recast for the purpose of VCSs, as defining canonically conjugate and continuous coordinates $(J_{\ell},\tau_{\ell})$, $\ell$ standing for the VCS index. Explicitly, if the Hamiltonian expectation value in any state can
be written as
\begin{equation}
\langle{\mathcal H}^{\rm red}\rangle=J_+\,\omega_+\ +\ J_-\,\omega_-
=\sum_\pm\,J_\pm\,\omega_\pm,
\end{equation}
where $\omega_{\pm}$ are some constant factors, we can identify through
the action-angle variational principle
\begin{equation}
\int\,dt\sum_\pm\,\left[\frac{d\tau_\pm}{dt}\,J_\pm\,-\,\omega_\pm\,J_\pm\right]\
\longleftrightarrow\
\int\,dt\left[\langle \frac{i}{\omega_0}\frac{d}{dt}\rangle\,-\,
\langle{\mathcal H}^{\rm red}\rangle\right]
\end{equation}
the following Hamiltonian equations
\begin{equation}
\frac{d\tau_\pm}{dt}
=\frac{\partial\langle{\mathcal H}^{\rm red}\rangle}{\partial J_\pm}
=\omega_\pm,\qquad
\frac{dJ_\pm}{dt}
=-\frac{\partial\langle{\mathcal H}^{\rm red}\rangle}{\partial\tau_\pm}
=0.
\label{varprin}
\end{equation}
Then any shift of time $\tau_{\pm}\to \tau_{\pm}+ t$ implies
that $\omega_{\pm}=1$ and, as a corollary, (\ref{varprin})
involves the canonical action coordinates conjugated to $\tau_{\pm}$
of the form
\begin{eqnarray}
J_+&=&|{\mathcal N}^+(|z|)|^2\,\cos^2\theta\,
\sum_{n=0}^\infty\, \frac{|z|^{2n}}{\left(K^0_+(\{n\})!\right)^2}
\left((h_{f}^{+}(n-1)!)h_{f}^{+}(0)\right)^2
\,e^+_{n},
\cr
J_-&=&|{\mathcal N}^-(|z|)|^2\,\sin^2\theta\,
\sum_{n=0}^\infty\, \frac{|z|^{2n}}{\left(K^0_-(\{n\})!\right)^2}\,
\left((h_{f}^{-}(n-1)!)h_{f}^{-}(0)\right)^2 \,
e^-_{n}.
\label{eq:acvar2}
\end{eqnarray}

\subsection{Explicit solutions}
\label{subsect43}
Here, we sketch the way to obtain
classes of solutions of the moment problems (\ref{eq:stj4})
whether in canonical or in deformed situations.

{\bf A simple class of solutions.}
We note that the general algebraic restriction such that
the ladder operators ${\mathcal M}^-$ and ${\mathcal M}^+$ obey
a $f$-deformed oscillator algebra on ${\mathcal V}$, namely
\begin{equation}
{\mathcal M}^-\,{\mathcal M}^+\,-\,{\mathcal M}^+\,{\mathcal M}^-
=\sum_{n=0,\pm}^\infty|e^\pm_{n}\rangle\,
(\{n+1\}-\{n\})\,\langle e^\pm_{n}|
\label{eq:const2}
\end{equation}
implies that $ K^0_\pm(\{n\})=\sqrt{\{n\}}$
with initial conditions $K^0_\pm(\{0\})=0$.
The normalization series (\ref{eq:fnorm2})
prove to be as ${\mathcal N}^{\pm}(|z|)=e^{-|z|^{2}/2}$ with infinite
convergence radii if one further sets
$(h_{f}^{\pm}(n))^{2}\,=\,(f(n+1))^{2}$.
The subsequent moment problems (\ref{eq:stj4}) reduce to
\begin{equation}
\int_0^{\infty}\,du\,u^n\,h^\pm(u)
= \,n !,\quad   n =0,1,2,\dots,
\label{eq:stj6}
\end{equation}
with solutions $h^{\pm}(u)=e^{-u}$
and associated weight factors as
${\mathcal W}^+\left(|z|\right)=3/(4\pi^2)$ and
${\mathcal W}^-\left(|z|\right)=3/(8\pi^2)$.

Hence, any family of $(k,\varepsilon,\kappa,f)$-VCSs possesses at
least one exact solution to their resolution of identity. Canonical
VCSs can be determined in the same way, with a constraint similar to
(\ref{eq:const2}), i.e. that the ladder operator should obey an
ordinary Fock-Heisenberg algebra.

{\bf A class of canonical solutions.}
The action identity constraint \cite{kl2} also bears
a class of solutions to the resolution of the identity equation
only for the canonical case $(f(N)\to 1, \kappa\to 1)$, a huge simplification
occurring in this case. Consider the requirements
\begin{equation}
J_+=\cos^2\theta \left(|z|^2 + e^+_{0}\right),
\qquad
J_-=\sin^2\theta \left(|z|^2 + e^-_{0}\right).
\label{eq:aic}
\end{equation}
These statements govern the next relations, under
a supplementary condition of a bounded from below energy spectrum such that
$e^{\pm}_{n}-e^{\pm}_{0}\geq 0$,
\begin{equation}
K_{\pm}^{0}(n)=\sqrt{e^{\pm}_{n}-e^{\pm}_{0}}.
\label{eq:solaic}
\end{equation}
The spin-orbit decoupled model with $\lambda(N)=0$ implies
\begin{equation}
\forall n\in \mathbb{N},\qquad
e^{\pm}_{n}-e^{\pm}_{0}=(1+\epsilon)n.
\end{equation}
One concludes that $K_{\pm}^{0}(n)=\sqrt{(1+\epsilon)n}$
and $K_{\pm}^{0}(0)=0$. The normalization factors of these nonlinear VCSs are $N^\pm(|z|)^{-2}=\exp[|z|^{2}/(1+\epsilon)]$ of
infinite convergence radii. The moment problem related to
the resolution of the identity of such states
can be put in the form
\begin{equation}
\int_{0}^{\infty}\;du \,u^{n}\,h^{\pm}(u)\,=\, (1+\epsilon)^{n}\,n!
\label{eq:stj3}
\end{equation}
Changing variables as $u\to u/(1+\epsilon)$,
it is not difficult to obtain
$h^{\pm}(u)=\exp[-u/(1+\epsilon)]/(1+\epsilon)$ and to deduce
the measure weight factors
\begin{equation}
{\mathcal W}^+\left(|z|\right)=\frac{3}{4\pi^2(1+\epsilon)},\qquad
{\mathcal W}^-\left(|z|\right)=\frac{3}{8\pi^2(1+\epsilon)}.
\label{eq:weight3}
\end{equation}

{\bf Solutions as $(p,q)$ deformations.}
We can also find classes of NVCSs with exact solutions
to their moment problem in $(p,q;\alpha,\beta,\ell)$-Burban deformed theory \cite{burban},
in which case
\begin{equation}
f(N)=\sqrt{\frac{p^{-\alpha\,N-\beta}-q^{\alpha\,N+\beta}}
{N(p^{-\ell}-q^{\ell})}},\quad
h_{f}^{\pm}(N)=\left(\frac{q^{\mu}}{p^{\nu}}\right)^{N}\sqrt{l^{\pm}(p,q)}
\label{eq:fpq-def}
\end{equation}
with $0<q<1$, $p>1$, $(pq)^\alpha <1 $, $\alpha\geq 0$, $\beta$,
$\ell$, $\mu$ and $\nu$ being real parameters and the positive real
valued functions $l^{\pm}(p,q)$ are such that $\lim_{(p,q)\to
(1^+,1^-)}h_f^{\pm}(N)=1$. Here, exact solutions to (\ref{eq:stj4})
can be expressed for $\beta=0$ and fixing the constant parameters
$(\mu,\nu,p,q,\alpha)$ so that the convergence radii of norm series
is nonvanishing. The problems (\ref{eq:stj4}) turn to
$(p,q;\alpha,0,\ell)$-Ramanujan integrals  of which solutions can be
written as deformed generalized exponential (see \cite{BGH2} and a
summary in the Appendix). Nevertheless, there is a more general
formulation recovering the $(p,q;\alpha,\beta,\ell)$-theory
and still allowing the existence of a class solutions that we
propose to investigate. The multi parameter $(p,q;\alpha,\beta,\ell;
\rho,\xi; \phi_1,\phi_2)$ quantum deformation is an extension of the
$(p,q;\rho,\xi; \phi_1,\phi_2)$-deformation as settled in Ref.
\cite{nk}, introducing the new indices $\alpha,\beta$ and $\ell$ as found
in (\ref{eq:fpq-def}). The deformation function in such a theory is
\begin{equation}
f(N)=\sqrt{\left(\frac{p^\rho}{q^\xi}\right)^N \frac{p^{-\alpha\,N-\beta}\phi_{1}(p,q)-q^{\alpha\,N+\beta}\phi_{2}(p,q)}
{N(p^{-\ell}-q^{\ell})}},
\label{eq:fpq-def2}
\end{equation}
with the set of conditions over parameters
\begin{eqnarray}
0 <\phi_1(p,q) < \phi_2(p,q), \quad \frac{\phi_1(p,q)}{\phi_2(p,q)}= (pq)^{k_0},\quad
k_0\in \mathbb{N},\\
(pq)^\alpha <1,\quad \alpha\geq 0, \quad p>1,\;\; 0<q<1, \;\;\;
(\beta,\ell) \in \mathbb{R}^2,
\end{eqnarray}
which ensures the convergence of the upcoming infinite sums and
products (see also \cite{nk} for relevant reductions). As a matter
of continuity, $\phi_i(p,q)$, $i=1,2$, could be taken continuous
functions of the two parameters $(p,q)$. As argued in \cite{nk},
recovered for $(\alpha,\beta,\ell)=(1,0,1)$, the integer $k_0$
causes the existence of two ground states $(k_0=0, k_0)$ of the
generalized harmonic oscillator Hamiltonian built in this framework.

Coming back to our problem, one imposes the next constraint on the
algebra of ladder operators
\begin{eqnarray}
&& \frac{q_0^\xi}{p_0^\rho}{\mathcal M}^-\,{\mathcal M}^+\,-q_0^\ell\,{\mathcal M}^+\,{\mathcal M}^-
=\sum_{n=0,\pm}^\infty|e^\pm_{n}\rangle\,
\frac{p_0^{(\rho-\alpha)\,N - \beta}}{q_0^{\xi \,N}}\,\phi_{1}(p_0,q_0)\,\langle e^\pm_{n}| \\
&&\frac{q_0^\xi}{p_0^\rho}{\mathcal M}^-\,{\mathcal M}^+\,-p_0^{-\ell}\,{\mathcal M}^+\,{\mathcal M}^-
=\sum_{n=0,\pm}^\infty|e^\pm_{n}\rangle\,
\frac{p_0^{\rho\,N}}{q_0^{(\xi-\alpha)\,N-\beta}}\,\phi_{2}(p_0,q_0)\,\langle e^\pm_{n}|,
\label{eq:cont3}
\end{eqnarray}
with $(p_0,q_0)$ the new deformation parameters such that $p_0>1$, $0<q_0<1$, $(p_0q_0)^\alpha<1$. A direct algebra shows that
\begin{eqnarray}
\frac{q_0^\xi}{p_0^\rho}\left(K^0_\pm([n+1]_0)\right)^2\,-\,q_0^{\ell}\,
\left(K^0_\pm([n]_0)\right)^2=\frac{p_0^{(\rho-\alpha)\,n - \beta}}{q_0^{\xi \,n}}\,\phi_{1}(p_0,q_0),\label{eq:fpqrecrel1}\\
\frac{q_0^\xi}{p_0^\rho}\left(K^0_\pm([n+1]_0)\right)^2\,-\,p^{-\ell}_0\,
\left(K^0_\pm([n]_0)\right)^2=\frac{p_0^{\rho\,n}}{q_0^{(\xi-\alpha)\,n-\beta}}\,\phi_{2}(p_0,q_0) ,
\label{eq:fpqrecrel2}
\end{eqnarray}
where
\begin{equation}
[n]_0=[n]_{(p_0,q_0)}:=\frac{p_0^{\rho\,n}}{q_0^{\xi\,n}}
\frac{(p^{-\alpha\,n-\beta}_0\,\phi_{1}(p_0,q_0)-q^{\alpha\, n+\beta}_0\,\phi_{2}(p_0,q_0))}{
(p^{-\ell}_0 - q_0^{\ell})}
\end{equation}
corresponds to basic integer of the theory.
The solutions to the recurrence relations (\ref{eq:fpqrecrel1})
and (\ref{eq:fpqrecrel2}) are
\begin{equation}
K^0_\pm([n]_0)=\sqrt{[n]_{(p_0,q_0)}}
\label{eq:sol1K0}
\end{equation}
with initial values $K^0_\pm([0]_0)=[0]_0$. Note that the symmetry
exchange $(p_0\leftrightarrow q^{-1}_0)$ $(\phi_1 \leftrightarrow
\phi_2)$ and $(\rho \leftrightarrow \xi)$,  makes compatible the
solutions of these recurrence relations. From the annihilation operator
action, one should set $K^0_\pm([0]_0)=[0]_0=0$ and  find a
restriction through
\begin{equation}
p^{-\beta}_0\,\phi_{1}(p_0,q_0)-q^{\beta}_0\,\phi_{2}(p_0,q_0)=0\quad \Longleftrightarrow
\quad \beta=k_0.
\end{equation}
This latter relation indicates that the annihilation operator action on the
ground state could vanish only when a Burban deformed theory
coincides with a $(p,q;\rho,\xi; \phi_1,\phi_2)$-theory up to a
deformation function, namely
\begin{eqnarray}
F(N) = \sqrt{\left(\frac{p^\rho}{q^\xi}\right)^N p^{-\beta}
\phi_{1}(p,q)\frac{(p^{-\ell}-q^{\ell})}{(p^{-1}-q^{1})} }.
\label{eq:equiv}
\end{eqnarray}
Moreover, in the former study \cite{BGH2}, the cancellation of
$K^0_\pm([0]_0)=[0]_0=0$ could be realized only for a linear theory
$\beta =0$. At this stage, noting that $k_0=\beta$ is still a free
parameter, the below solutions prove to be more general than ones
obtained in that latter reference.

The norm series of the NVCSs become, using the expression of
$h^\pm_f(N)$ (\ref{eq:fpq-def}),
\begin{equation}
|{\mathcal N}^\pm(|z|)|^{-2}=
\sum_{n=0}^\infty\,\frac{|z|^{2n}}{[n]_{0}!}
\left(\frac{q^{\mu}}{p^{\nu}}\right)^{n(n-1)}\,(l^{\pm}(p,q))^{n}\,,
\end{equation}
and have the radius of convergence
\begin{eqnarray}
&&R_\pm=\lim_{n\to\infty}
\left[\left(\frac{q^{\mu}}{p^{\nu}}\right)^{-2n}
\frac{[n]_0}{l^{\pm}(p,q)}\right]^{1/2}\\
&& =\lim_{n\to\infty}
\left[\left(p_0^{\rho-\alpha}\,p^{2\nu}q^{-2\mu}q_0^{-\xi}\right)^{n}\, p_0^{-\beta}\phi_1(p,q)
\frac{1-(p_0q_0)^{\,\alpha\, n}}{l^{\pm}(p,q)(p_0^{-\ell}- q_0^\ell))}\right]^{1/2}.
\end{eqnarray}
Assuming that $p_0^{\rho-\alpha}\,p^{2\nu}q^{-2\mu}q_0^{-\xi}>1$,
$R_\pm$ are infinite. The moment problem (\ref{eq:stj4}) can be written as,
for all $n\in\mathbb{N}$,
\begin{eqnarray}
&& \int_0^\infty\,du\,u^n\,h^\pm(u)=
\left(\frac{q^{\mu}}{p^{\nu}}\right)^{-n(n-1)}\,
\left(l^{\pm}(p,q)\right)^{-n} \; [n]_{0}!\cr
&& \ \ \ \ \ = q_0^{-\xi \frac{n(n+1)}{2}}q^{-\mu n(n-1)}p_0^{\rho \frac{n(n+1)}{2}-\beta n}
p^{-\nu n(n-1)} \Psi(p,q)^n \; [p^\alpha, q^\alpha; p^\alpha, q^\alpha]_n,\\\cr
&&\Psi(p,q) = \phi_1(p,q) (p_0^{-\ell}- q_0^\ell)^{-1} (l^{\pm}(p,q))^{-1}.
\label{eq:stj7}
\end{eqnarray}
We now choose an appropriate set of parameters as
\begin{equation}
p_0 = p,\;\;\; q_0 = q,\qquad \frac{\xi}{2}+\mu = \frac{\alpha}{2},\qquad
 \frac{\rho}{2}+\nu =0,
\end{equation}
so that $p_0^{\rho-\alpha}\,p^{2\nu}q^{-2\mu}q_0^{-\xi}=(pq)^{-\alpha}>1$,
implying infinite radii of convergence for the norm series.
Afterwards, making use of the $(p^\alpha,q^\alpha)$-extension of the Ramanujan integral, we derive the moment function
(generalized exponential functions can be found in the Appendix)
\begin{eqnarray}
&& h^\pm\left(|z|^2\right)= \Phi^{-1}(p,q) \frac{1}{\log(1/(pq)^\alpha)}
e_{(p^\alpha,q^\alpha)} \left(-|z|^2\,\Phi^{-1}(p,q) p^{-\alpha/2}\right), \cr
&&\Phi(p,q) = q^{\alpha - \xi} p^{-2\nu- \beta}\Psi (p,q).
\label{eq:sol-mom}
\end{eqnarray}
The norm series can be inferred as,
$\left(K^0_\pm([n])\right)^2=[n]$,
\begin{equation}
|{\mathcal N}^\pm(|z|)|^{-2} ={\mathcal
E}^{(1/2,0)}_{(p^{\alpha},q^{\alpha})} \left(|z|^2
q^{\xi-\alpha/2}p^{\beta +2\nu}\phi_1^{-1}(p,q)l^{\pm}(p,q)
\left(p^{-\ell}-q^\ell\right)\right),
\end{equation}
and one can easily deduce the weight functions ${\mathcal W}^\pm(|z|)$
from (\ref{eq:hfunctions}). These weights generalize the measure as obtained in
\cite{BGH2} with a freedom parametrized by the deformation functions
$\phi_1(p,q)$, and the new extra parameters $q^{\xi}$ and $p^{\beta +2\nu}$.
In order to get previous results of \cite{BGH2}, one has to set
$\xi=0=\beta +2\nu$ and $\phi_1(p,q)=1$ in which case $\phi_{2}(p,q)$
corresponds to the monomial function $(pq)^{-\beta}$.

In summary of this study, NVCSs associated with nonlinear spin-orbit Hamiltonian
are characterized by a unit vector of the sphere $S^2$ determined by
coordinates $\theta$ and $\phi$. Some appropriate constraints should be
set in order that they could fulfill all the axioms of Gazeau-Klauder, namely continuity in the parameter $z\in \mathbb{C}$, temporal stability through a shift of the real parameters $\tau_\pm \to  \tau_\pm + t$ and the overcompleteness property
as a resolution of the Hilbert space ${\mathcal V}$.
They distinguish from one another by different real positive factors
$K^0_\pm(\{n\})$ parameterizing the freedom afforded by the annihilation operator action.
An exact resolution of the unity over the total Hilbert space
can be derived after specifying the remaining freedom.

\section{Matrix formulation}
\label{sect5}

The $S^2$ NVCSs have a natural extension as matrix NVCSs.
In Ref. \cite{BGH2}, we considered diagonal and quaternionic matrix
domains. Here, we enlarge the study to a normal (including
diagonal complex matrices) and quaternionic matrix domains
taking of course into account the new Hilbert space framework.

To proceed, we write the diagonal matrix annihilation operator
associated with (\ref{diagan2}) as
\begin{eqnarray}
\fl \mathbb{M}^{-}= \sum_{n=0,\pm}^\infty\,|n-1\rangle\,\langle n|\otimes
K(\{n\}) \, |\pm\rangle\,\langle \pm|, \quad
K(\{n\})= {\rm diag}(K_+(\{n\}), K_-(\{n\}))
\label{matann}
\end{eqnarray}
where we assume that $K(\{n\})$ is in diagonal form.
By similarity, we project the annihilation operator onto the basis $|e^\pm_n\rangle$ as
\begin{eqnarray}
{\mathcal M}^{-}={\mathcal U}\,\mathbb{M}^{-}\,{\mathcal U}^\dagger=
\sum_{n=0,\pm}^\infty\, K(\{n\}) |e_{n-1}^\pm\rangle\,\langle e_n^\pm|.
\end{eqnarray}
The same eigenvalue problem as in (\ref{eq:ev1}) is found  as
\begin{eqnarray}
{\mathcal M}^-|(z,w);\tau_\pm;\pm\rangle =
\widetilde{\mathfrak{Z}}(z,w)\,\widetilde{\mathbb{Q}}_{\mathcal V}\,
|(z,w);\tau_\pm;\pm\rangle,
\label{eq:eigen}
\end{eqnarray}
with the same shape for the operators $\widetilde{\mathfrak{Z}}(z,w)$ and $\widetilde{\mathbb{Q}}_{\mathcal V}$.
Still assuming that $\mathfrak{Z}={\mathcal U}^\dagger\;\widetilde{\mathfrak{Z}} \;{\mathcal U}$
is a complex constant matrix, we define
$\mathbb{Q}_{\mathcal V}:={\mathcal U}^\dagger \widetilde{\mathbb{Q}}_{\mathcal V} \;{\mathcal U}$,
so that (\ref{eq:eigen}) can be put into the form
\begin{eqnarray}
\mathbb{M}^-|\mathfrak{Z};\tau_\pm;\pm\rangle =
\mathfrak{Z}\,\mathbb{Q}_{\mathcal V}\,
|\mathfrak{Z};\tau_\pm;\pm\rangle,\quad
|\mathfrak{Z};\tau_\pm;\pm\rangle = {\mathcal U}^\dagger\;
|(z,w);\tau_\pm;\pm\rangle.
\label{eq:eigen2}
\end{eqnarray}
Let us assume that $\mathbb{Q}_{\mathcal V}$ admits the expansion
\begin{eqnarray}
\mathbb{Q}^\mathfrak{U}_{\mathcal V}=
\sum_{n=0,\pm}^\infty\, |n\rangle\,\langle n|\otimes \mathfrak{U}\, h_f(n)\,  \mathfrak{U}^\dag|\pm\rangle\,\langle \pm| ,
\quad h_f(n)={\rm diag}(h_f^+(n),h_f^-(n)) ,
\nonumber
\end{eqnarray}
with $\mathfrak{U}$ an element of the unitary group $U(2)$ or an
element of $SU(2)$. Let us fix ${\mathfrak{Z}}$ as a normal matrix
and ${\mathfrak{Z}}_{\rm quat}$ as a quaternionic matrix, i.e.,
they satisfy
\begin{eqnarray}
&&{\mathfrak{Z}}^\dag {\mathfrak{Z}} = {\mathfrak{Z}} {\mathfrak{Z}}^\dag,\quad
\;\;\; {\mathfrak{Z}} = V\, {\rm diag}(z, w) \,V^\dag, \quad V \in U(2),\\
&& \mathfrak{Z}_{\rm quat} =U \,{\rm diag}(z,\bar z) \,U^\dag,\quad U \in SU(2),\label{quater}\\
&& U = u_{\phi_1} u_\theta  u_{\phi_2},  \;\;
u_{\theta}= \left(\begin{array}{cc}
\cos \frac{\theta}{2} & i \sin \frac{\theta}{2} \\
 i \sin \frac{\theta}{2} &\cos \frac{\theta}{2}
\end{array}\right) ,\;\;
u_{\phi_i}= \left(\begin{array}{cc}
e^{i \frac{\phi_i}{2}} & 0\\
 0 &e^{- i \frac{\phi_i}{2}}
\end{array}\right),
\end{eqnarray}
for $\theta\in [0,\pi]$ and $\phi_i\in ]0, 2\pi]$.
Therefore, the operators
\begin{eqnarray}
&&\mathfrak{Z}_f:= \mathfrak{Z}\,\mathbb{Q}^{V}_{\mathcal V}\, = \sum_{n=0,\pm}^\infty\, |n\rangle\,\langle n|\otimes V\,\mathfrak{Z}_{\rm diag} h_f(n)\,  V^\dag|\pm\rangle\,\langle \pm|,\\
&&\mathfrak{Z}_{\rm diag} :={\rm diag}(z,w)\\
&& \mathfrak{Z}_{{\rm quat},f}:= \mathfrak{Z}_{\rm quat}\,\mathbb{Q}^U_{\mathcal V}= \sum_{n=0,\pm}^\infty\, |n\rangle\,\langle n|\otimes U\,\mathfrak{Z}_{\rm qdiag} h_f(n)\,  U^\dag|\pm\rangle\,\langle \pm|,\\
&&\mathfrak{Z}_{\rm qdiag} :={\rm diag}(z,\bar z),
\end{eqnarray}
satisfy the equations
\begin{eqnarray}
\mathbb{M}^-_V|\mathfrak{Z};\tau_\pm;\pm\rangle =
\mathfrak{Z}_f |\mathfrak{Z};\tau_\pm;\pm\rangle,
\label{eq:eigeva1}\\
\mathbb{M}^-_U|\mathfrak{Z}_{\rm quat};\tau_\pm;\pm\rangle =
\mathfrak{Z}_{{\rm quat},f} |\mathfrak{Z}_{\rm quat};\tau_\pm;\pm\rangle,
\label{eq:eigeva2}
\end{eqnarray}
obtained by substituting in (\ref{eq:eigen2}), $\mathbb{Q}_{\mathcal V}$ by
$\mathbb{Q}^\mathfrak{U}_{\mathcal V}$, for $\mathfrak{U}=V,U$,
and using the new group dependent annihilation operators
\begin{eqnarray}
\mathbb{M}^-_{\mathfrak{U}}=
\sum_{n=0,\pm}^\infty\,|n-1\rangle\,\langle n|\otimes
\mathfrak{U}\, K(\{n\}) \,\mathfrak{U}^\dag\, |\pm\rangle\,\langle \pm|,\qquad
\mathfrak{U}=V,U.
\end{eqnarray}

The procedure as applied in Subsection \ref{subsect41} can be used
for obtaining the general solutions of the eigenvalue problem (\ref{eq:eigeva1})
and (\ref{eq:eigeva2}) with all Gazeau-Klauder properties.
We have the general form of Gazeau-Klauder NVCSs:

$\bullet$ For normal matrices
\begin{eqnarray}
|\mathfrak{Z};\tau_\pm;\pm\rangle
&=&N(\mathfrak{Z})\sum_{n=0}^{\infty}\, |n\rangle  \otimes V\,
(K(\{n\})!)^{-1}\, V^\dag \; \mathfrak{Z}_f^{n}\,|\pm\rangle
\label{eq:matvcs}\\
&=&N(\mathfrak{Z})\sum_{n=0}^{\infty}\, |n\rangle  \otimes
V \, R_0(n)\exp[-i\omega_0\tau e_{n}]\;
\mathfrak{Z}_{\rm diag}^{n}\,  V^\dag \,|\pm\rangle,
\label{eq:matvcs2}\\
N(\mathfrak{Z})^{-2}&=&
\sum_{n=0}^{\infty}\,{\rm Tr}\left[ V \, |R_0(n)|^2\, {\rm diag}(|z|^{2n}, |w|^{2n}) V^\dag \right]\cr
&=&\sum_{n=0}^{\infty}\,
\left(|z|^{2n} \,|R^0_{+}(n)|^2 + |w|^{2n} \,|R^0_-(n)|^2\right),
\label{eq:norma}\\
R^0(n)&=&{\rm diag}(R^0_+(n) ,R^0_-(n))
=(K^0(\{n\})!)^{-1}(h_f(n-1)!h_f(0)),
\end{eqnarray}
where $K^0(\{n\})= {\rm diag}(K_+^0(\{n\}), K^0_-(\{n\}))$,
$K_\pm^0(\{n\})=|K_\pm(\{n\})|$, $N(\mathfrak{Z})$ is the normalization factor,
$\tau={\rm diag}(\tau_+,\tau_-)$ and
$e_n={\rm diag}(e^+_{n},e^-_{n})$.
Note that the convergence radii of the series (\ref{eq:norma})
are such that $|z|\leq L_+$, $|w|\leq L_-$ and
$L_\pm=\lim_{n\to\infty} K^0_\pm(\{n\})/|h_{\pm f}(n-1)|$.

$\bullet$ For quaternion matrices
\begin{eqnarray}
&&|\mathfrak{Z}_{\rm quat};\tau_\pm;\pm\rangle
=
N(\mathfrak{Z}_{\rm quat})\sum_{n=0}^{\infty}\, |n\rangle  \otimes
U\, (K(\{n\})!)^{-1}\, U^\dag\; \mathfrak{Z}_{{\rm quat},f }^{n}\,|\pm\rangle\\
&&\ \ \ \ \ \ \ \ \ \ \ \ \ \
= N(\mathfrak{Z}_{\rm quat})\sum_{n=0}^{\infty}\, |n\rangle  \otimes
U \, R_0(n)\exp[-i\omega_0\tau e_{n}]\,
\mathfrak{Z}_{\rm qdiag}^{n} \, U^\dag \, |\pm\rangle,
\label{eq:quatvcs}\\
&&N(\mathfrak{Z}_{\rm quat})^{-2}=
\sum_{n=0}^{\infty}\,{\rm Tr}\left[U  \, |R_0(n)|^2\,|z|^{2n} \mathbb{I}_2\, U^\dag \right]\cr
&&\ \ \ \ \ \ \ \ \ \ \ \ \ \
=\sum_{n=0}^{\infty}\,
|z|^{2n}\left(|R^0_{+}(n)|^2+|R^0_-(n)|^2\right),
\label{eq:normqua}
\end{eqnarray}
with the norm series convergence radius
\begin{eqnarray}
L = \lim_{n\to\infty} \left[ \frac{|R^0_{+}(n)|^2+|R^0_-(n)|^2}{|R^0_{+}(n+1)|^2+|R^0_-(n+1)|^2}\right]^{\frac{1}{2}}.
\end{eqnarray}
The NVCSs (\ref{eq:matvcs2}) and (\ref{eq:quatvcs}) are continuous,
normalized according to the definition
\begin{eqnarray}
\sum_{\pm} \langle \mathcal{Z};\tau_\pm;\pm|\mathcal{Z};\tau_\pm;\pm\rangle =1 ,\qquad \mathcal{Z} = \mathfrak{Z},\;\; \mathfrak{Z}_{\rm quat},
\end{eqnarray}
and stable under the time evolution operator
\begin{eqnarray}
U^{\mathfrak{U}}(t) =
 \exp[-i \omega_0 t\, \mathfrak{U}\,\mathbb{H}^{\rm red}\,\mathfrak{U}^\dag]= \mathfrak{U}\, \exp[-i \omega_0 t \mathbb{H}^{\rm red}]\,\mathfrak{U}^\dag,
 \quad \mathfrak{U} = V,U,
\end{eqnarray}
i.e.
\begin{eqnarray}
U^{\mathfrak{U}}(t) |\mathcal{Z};\tau_\pm;\pm\rangle = |\mathcal{Z};\tau_\pm+t;\pm\rangle.
\end{eqnarray}
The action identity axiom could also be inferred by assigning the
action variables to
\begin{eqnarray}
J^{\mathfrak{U}}_\pm = \langle \mathcal{Z};\tau_\pm;\pm |\mathfrak{U}\, \mathbb{H}^{\rm red}  \,\mathfrak{U}^\dag|\mathcal{Z};\tau_\pm;\pm  \rangle,
\qquad \mathfrak{U} = V,U.
\end{eqnarray}
To recover the original NVCSs over the basis $|e_n^\pm\rangle$,
we derive the state (\ref{eq:eigen2}) as follows
\begin{eqnarray}
|(z,w);\tau_\pm;\pm\rangle_{\mathcal{Z}} & =&  \mathfrak{U}\, {\mathcal U}\, \mathfrak{U}^\dag\,|\mathcal{Z};\tau_\pm;\pm\rangle \\
&= &N(\mathcal{Z})\sum_{n=0}^{\infty}\,   \mathfrak{U}\,|e_n^\pm \rangle
\, R_0^\pm(n)\exp[-i\omega_0\tau_\pm e^\pm_{n}]\;\mathcal{Z}^{n}_\pm\,\mathfrak{U}^\dag_\pm,
\end{eqnarray}
with $\mathcal{Z}_\pm = \langle\pm |\mathcal{Z}_{\rm diag}|\pm\rangle$
and $\mathfrak{U}^\dag_\pm = \langle\pm |\mathfrak{U}^\dag|\pm\rangle$, so that
the operator $\widetilde{\mathfrak{Z}}$ has the form
\begin{equation}
 \widetilde{\mathfrak{Z}}_{\mathcal{Z}} =
\sum_{n=0,\pm}^{\infty}\,   \mathfrak{U}\;| e_n^\pm \rangle \mathcal{Z}_\pm  \langle e_n^\pm|\;\mathfrak{U}^\dag\,.
\label{eq:eig}
\end{equation}
One notes that the normalization factor of $|(z,w);\tau_\pm;\pm\rangle$ remains
the same as that of $|\mathcal{Z};\tau_\pm;\pm\rangle$ as expected from any
unitary transformation.

First, let us treat the resolution of the identity related to normal matrix domain.
The overcompleteness relation of the normal matrix NVCSs is given by
\begin{equation}
\mathbb{I}_{\mathcal V} = \sum_\pm \int_{\mathcal D} d\mu(\mathfrak{Z})
|\mathfrak{Z};\tau_\pm;\pm\rangle\, \langle \mathfrak{Z};\tau_\pm;\pm |,
\label{resolut}
\end{equation}
where the domain ${\mathcal D}$ and the measure $d\mu(\mathfrak{Z})
$ are to be defined. Considering the parametrization of the variable
$\mathfrak{Z} = V\, {\rm diag}(z,w)\, V^\dag$ as
\begin{eqnarray}
&& z = r_+ e^{i \theta_+},\;\;  w = r_- e^{i \theta_-}, \quad r_\pm \in [0,L_\pm),
\;\;\;
\theta_\pm \in [0,2\pi [,
\end{eqnarray}
then the domain of integration ${\mathcal D}$ is given by
\begin{eqnarray}
{\mathcal D} =  [0,L_+)\times [0,L_-)\times \{[0,2\pi [\}^{ 2}
\times U(2),
\end{eqnarray}
where one should include the Lie group $U(2)$.
Therefore the measure $d\mu(\mathfrak{Z})$ is of the form
\begin{eqnarray}
d\mu(\mathfrak{Z}) = N(\mathfrak{Z})^{-2}\,{\mathcal W}_+(r_+)\,{\mathcal W}_-(r_-)\,
r_+\,r_-\,dr_+\;dr_-\; d\theta_+\;d\theta_- d\Omega_{U(2)}(V);
\label{normes}
\end{eqnarray}
here $d\Omega_{U(2)}(V)$ is the invariant Haar measure over $U(2)$ normalized to one,
${\mathcal W}_\pm(r_\pm)$ are weight factors to be fixed later.

After integration over the angle variables $\theta_\pm$, the
identity (\ref{resolut}) involves the next integral over the group
$U(2)$,
\begin{eqnarray}
&&\fl  \int_{U(2)} d\Omega_{U(2)}(V)\;  V\,
((R_+^0(n))^2 r_+^{2n}|+\rangle\, \langle +| + (R_-^0(n))^2 r_-^{2n}|-\rangle\, \langle -|) \; V^\dag
\cr
&&\fl \ \ \ \ \ \ \ =
\frac{1}{2} r_+^{2n} (R^0(n)_+)^2 + \frac{1}{2} r_-^{2n} (R^0(n)_-)^2,
\end{eqnarray}
where we have used the orthogonality condition on compact group \cite{al}
\begin{equation}
\int_{U(2)}\; d\Omega_{U(2)}(V)\;\; V\;|\pm\rangle\, \langle \pm| \;
V^\dag = \frac{1}{2} \mathbb{I}_2;
\end{equation}
here $\{|\pm\rangle\}$ plays the role of the canonical basis
of $\mathbb{C}^2$. We end with the moment problems
\begin{equation}
\int_0^{L^2_\pm} d u_\pm\, u_\pm^n  h_\pm(u_\pm)  =(K^0_\pm(\{n\})!)^2/ (h_{\pm f}(n-1)!h_{\pm f}(0))^2,
\label{eqmom}
\end{equation}
where $u_\pm= r^2_\pm$ and the functions $h_\pm(u_\pm)
= \pi {\mathcal W}_{\pm}(u_\pm)$.

Solution to (\ref{eqmom}) could be found along the lines of
Subsection \ref{subsect43} for canonical and deformed situation.
We give here the simplest solution obtained by stressing the ladder
operators to satisfy $[\mathbb{M}^-,\mathbb{M}^+ ]=
(\{N+1\}-\{N\})\mathbb{I}_2$, implying $K_{\pm}(\{n\}) = \{n\}$ and
fixing $(h_{\pm f}(n))^2=(f(n+1))^2$. The norm series
(\ref{eq:norma}) are $N(\mathfrak{Z})^{-2}=(e^{r_+^2} +e^{r_-^2})$
such that $L_\pm =\infty$. One can solve (\ref{eqmom}) and find  $
h_\pm(u_\pm)=e^{-u_\pm}$ from which the weight factors ${\mathcal
W}_\pm(u_\pm)$ and the measure
\begin{equation}
d\mu(\mathfrak{Z}) = \frac{1}{\pi^2}\, (e^{-r_+^2}+e^{-r_-^2})\,
r_+\,r_-\,dr_+\;dr_-\; d\theta_+\;d\theta_- d\Omega_{U(2)}(V)
\label{mes0}
\end{equation}
are deduced. For other deformed theories, one can also show  that
these moment problems find solutions using appropriate deformed
exponential functions. Last, the correct measure for the anterior
$\widetilde{\mathfrak Z}$-NVCSs can be easily obtained by using the
fact that $d\mu(\mathfrak{Z})$ is an invariant measure.

Concerning the quaternionic matrix domain, a similar treatment can
be applied. However, we use a different technique in order to avoid
the integration over the Lie group. The measure will be endowed with
the following parametrization of the quaternions $\mathfrak{Z}_{\rm
quat} = U\, {\rm diag}(z, \bar z)\, U^\dag$ as, using direct
expansion of (\ref{quater}),
\begin{eqnarray}
\mathfrak{Z}_{\rm quat} = r (\cos \xi\, \mathbb{I}_2 + i \sin \xi \sigma),
\qquad
\sigma = \left(\begin{array}{cc}
\cos \theta & e^{i \phi} \sin\theta\\
 e^{- i \phi} \sin\theta & -\cos \theta
\end{array}\right)
\end{eqnarray}
where
\begin{eqnarray}
z = r e^{i \xi}, \quad  r  \in [0,L),\;\;\; \xi \in [0,2\pi [,\;\;\;
\theta \in [0,\pi[,\;\;\; \phi \in [0,2\pi[.
\end{eqnarray}
Note that the Lie group $SU(2)$ dependence has been traded for a $S^2$ unit vector
indices. It is then crucial to observe that, since $\sigma^2 =\mathbb{I}_2$,
\begin{eqnarray}
\mathfrak{Z}_{\rm quat} = r \exp[ i\, \xi \, \sigma],
\end{eqnarray}
thence any power of $\mathfrak{Z}_{\rm quat}$ can be easily deduced
and $(\mathfrak{Z}_{\rm quat})^\dag =  r \exp[- i\, \xi \, \sigma]$.

The domain of integration ${\mathcal D}$ is nothing but
\begin{eqnarray}
{\mathcal D} =  [0,L)\times [0,2\pi[ \times S^2
\end{eqnarray}
for which the measure $d\mu(\mathfrak{Z}_{\rm quat})$ appears as
\begin{eqnarray}
\fl d\mu(\mathfrak{Z}_{\rm quat}) = N(\mathfrak{Z})^{-2}\,{\mathcal
W}(r)\, r dr \, d\xi\, d\mu_{S^2}(\theta,\phi),\quad
d\mu_{S^2}(\theta,\phi) = \frac{1}{4\pi} \, \sin\theta d\theta
\,d\phi \label{qames}
\end{eqnarray}
with ${\mathcal W}(r)$ a weight factor to be determined. A
straightforward algebra induces the moment problem
\begin{equation}
\int_0^{L^2} d u\, u^n  h(u)  =(K^0_\pm(\{n\})!)^2/ (h_{\pm f}(n-1)!h_{\pm f}(0))^2,
\label{eqmom2}
\end{equation}
where $u= r^2$ and the functions $h(u)= \pi {\mathcal W}(u)$. Again,
we can single out a solution of the problem (\ref{eqmom2}) by
constraining the function $(h^0_{+ f}(n))^2=(h^0_{-
f}(n))^2=(f(n+1))^2$, and the algebra to be such that
$[\mathbb{M}^-,\mathbb{M}^+ ]= (\{N+1\}-\{N\})\mathbb{I}_2$ implying
$K_{\pm}(\{n\}) = \{n\}$. Indeed, we find the norm series
(\ref{eq:normqua}) is $N(\mathfrak{Z}_{\rm quat})^{-2}= (2 e^{r^2})$
with convergence radius $L =\infty$. Therefore the solution of
(\ref{eqmom2}) as $ h(u)=e^{-u}$ corresponds to a measure
\begin{equation}
d\mu(\mathfrak{Z}_{\rm quat}) = \frac{1}{2\pi^2}\,
r\,dr\; d \xi\; d\theta\;d\phi\; d\mu_{S^2}.
\label{mes1}
\end{equation}
The measures (\ref{mes0}) and (\ref{mes1}) have  been discussed in
Refs \cite{thirolu,BGH2}. Finally, the same comments about solvable
deformed theories as over normal NVCSs remain true also for
quaternion NVCSs.

\section{Deformed displacement operators, dual states and $T-$operators}
\label{sect6}

Unitary displacement operators for VCSs over matrix domain
have a well defined sense via tensor product of matrix and Hilbert spaces \cite{al}.
Here, we need to define deformed versions of these operators
but still generating the $S^2$ or matrix NVCSs. A step forward
is to regard the displacement operators in the context
of NCSs \cite{roy} which are mainly exploited
in deformed quantum optics \cite{nad,rok,rok2}.
In Ref. \cite{tavas}, deformed inverse bosonic operators are used
and, from these, their dual NCSs are defined.
The ensuing purpose is the extension of these operators
to deformed versions of displacement operators and to investigate
how they generate $S^2$ and matrix NVCSs.

\subsection{ $S^2$-displacement operators}

Consider the annihilation operator
\begin{eqnarray}
{\mathcal M}^{-} =\sum_{n=0,\pm}^{\infty}\,| e^{\pm}_{n-1}\rangle\;
K_{\pm}(\{n\})\;\langle e^{\pm}_{n}|,
\label{eq:ann21}
\end{eqnarray}
where $K_{\pm}(\{n\})$ are free functions of $\{n\}$.
Let us define a new operator
\begin{eqnarray}
{\mathcal B}^{+} =\sum_{n=0,\pm}^{\infty}\,| e^{\pm}_{n+1}\rangle\;
\overline{G_{\pm}(\{n+1\})}\;\langle e^{\pm}_{n}|,
\label{eq:ann22}
\end{eqnarray}
where $\overline{G_{\pm}(\{n\})}$ are new functions of the number
$\{n\}$ and impose the condition
\begin{eqnarray}
[ {\mathcal M}^{-}, {\mathcal B}^{+}] = \mathbb{I}_{\mathcal V}.
\label{adj}
\end{eqnarray}
Therefore, a direct evaluation of (\ref{adj}) proves that
 the pair $(K_{\pm}(\{n\}),\overline{G_{\pm}(\{n\}))}$ should satisfy
\begin{eqnarray}
&&K_{\pm}(\{1\}) \overline{G_{\pm}(\{1\})}= 1,\\
&&K_{\pm}(\{p+1\}) \overline{G_{\pm}(\{p+1\})}
- K_{\pm}(\{p\}) \overline{G_{\pm}(\{p\})}=1,\;\;\;
\forall p\geq 1.
\end{eqnarray}
A simple solution of this problem is
\begin{eqnarray}
G_{\pm}(\{0\})= 0, \qquad \overline{G_{\pm}(\{p\})} = p\, (K_{\pm}(\{p\}))^{-1},
\;\;\forall p\geq 1.
\end{eqnarray}
Moreover, one can check that
\begin{eqnarray}
[ \widetilde{\mathbb Q}_{\mathcal V}^{-1}{\mathcal M}^{-}, {\mathcal B}^{+}\widetilde{\mathbb Q}_{\mathcal V}] = \mathbb{I}_{\mathcal V},\qquad
\widetilde{\mathbb{Q}}^{-1}_{\mathcal V}=
\sum_{n=0,\pm}^\infty\,|e^{\pm}_{n}\rangle\,
(h_{f}^{\pm}(n))^{-1}\,\langle e^{\pm}_{n}|.
\label{adj2}
\end{eqnarray}
We are immediately in position to define a displacement operator
for $S^2$ NVCS theory by introducing
\begin{eqnarray}
{\mathcal D}_f = e^{z \, {\mathcal B}^{+}\widetilde{\mathbb Q}_{\mathcal V}
-\bar z \, \widetilde{\mathbb Q}_{\mathcal V}^{-1}{\mathcal M}^{-} }
\end{eqnarray}
and the NVCSs can be rebuilt from
the action of the operator ${\mathcal D}_f$  onto the ground states as
\begin{eqnarray}
&&\fl |z;\tau_\pm;\theta,\phi\rangle =
{\mathcal D}_f \left\{  e^{\frac{1}{2}|z|^2}
\left[ {\mathcal N}^+(|z|)\,\cos\theta\,e^{-i\omega \tau_+ e^+_0 }|e^+_{0}\rangle
 +  {\mathcal N}^-(|z|)\,e^{i\phi}\,\sin\theta\,
 e^{-i\omega \tau_- e^-_0 }|e^-_{0}\rangle \right] \right\}.\cr
&&\label{eq:fcoh212}
\end{eqnarray}
Of course, the limit $f(N)\to 1$ implies that ${\mathcal D}_f$
converges to a kind of ordinary displacement operator $e^{za^\dag -
\bar z a}$ of usual CSs recovered for $K(n)= \sqrt{n}$.

Another class of NCVSs, so-called ``dual'' NVCSs, could be
introduced by noting the fact that
\begin{eqnarray}
&&[ {\mathcal M}^{+}, {\mathcal B}^{-}] = \mathbb{I}_{\mathcal V},
\qquad
[\widetilde{\mathbb Q}_{\mathcal V}{\mathcal B}^{-},
{\mathcal M}^{+}\widetilde{\mathbb Q}_{\mathcal V}^{-1}]  = \mathbb{I}_{\mathcal V},
\qquad {\mathcal B}^{-} =  ({\mathcal B}^{+})^\dag,
\end{eqnarray}
and by defining a new displacement operator of the form
\begin{eqnarray}
{\mathcal D}'_f = e^{z \, {\mathcal M}^{+}\widetilde{\mathbb Q}_{\mathcal V}^{-1}
-\bar z \, \widetilde{\mathbb Q}_{\mathcal V}{\mathcal B}^{-} }.
\end{eqnarray}
Applying the latter on ground states, we get the set of dual NVCSs as
\begin{eqnarray}
&&
\fl
 |z;\tau_\pm;\theta,\phi\rangle' =
{\mathcal D}'_f \left\{  e^{\frac{1}{2}|z|^2}
\left[  {\mathcal N}'^+(|z|)\cos\theta\,e^{+i\omega \tau_+ e^+_0 }|e^+_{0}\rangle
 +  {\mathcal N}'^-(|z|)\,e^{i\phi}\sin\theta
 e^{+i\omega \tau_- e^-_0 }|e^-_{0}\rangle \right] \right\}\cr
 &&
\label{eq:fcoh213}
\end{eqnarray}
where one needs to introduce new normalization
factors
\begin{equation}
{\mathcal N}'^\pm(|z|)=
\left[
\sum_{n=0}^\infty\,
\frac{|z|^{2n}}{(n!)^2}\frac{\left(K^0_\pm(\{n\})!\right)^2}
{\left((h_{f}^{\pm}(n-1)!)h_{f}^{\pm}(0)\right)^{2}}
\right]^{-1/2}
\label{eq:fnorm3}
\end{equation}
of convergence radii $R'_\pm$
\begin{equation}
R'_\pm=\lim_{n\to\infty}\left[
\frac{n\,h_{f}^{\pm}(n-1)}{K^0_\pm(\{n\})}\right].
\label{eq:fradii2}
\end{equation}
Noting the change of phase $+i\omega \tau_\pm e^\pm_0$ (\ref{eq:fcoh213}),
a peculiar notion of evolution is encoded in the dual state definition.
We will discuss this hereunder according to the matrix formulation.

One could ask if the NVCSs and their analogue dual states could be
simultaneously solvable in the sense that they give an exact resolution
of the identity. Let us consider the specific instance when
$K^0_\pm(\{n\})=\sqrt{\{n\}}$ and $h_f^\pm(n)= sf(n+1)$, $s=\pm$,
since this case proves to be solvable for NVCSs (See Subsection
\ref{subsect43}). Therefore the resolution of the identity for
dual NVCSs reduces to the moment problem (\ref{eq:stj6}), for $s=+$,
$R'=\lim_{n\to\infty}\sqrt{n}=\infty$ and then getting
$h_\pm(u_\pm)=e^{-u_\pm}$. These solutions consist of those of any
set of CSs coinciding with their dual counterpart in the absence of
deformation, i.e., $f(N)\to 1$ \cite{rok2}. For other deformed
theories, any answer could be given without making a careful
analysis. However, disregarding the prime NVCSs, we can always find
an exact  resolution of the moment problem for $(p,q)$ deformation
of the dual NVCSs. Indeed, one has to switch the role played by
$K(\{n\})$ and $h_f^\pm(n-1)$ along the lines of the resolution in
Subsection \ref{subsect43} and then totally finds  similar
solutions. Another remarkable feature introduced by the operator
${\mathbb Q}_{\mathcal V}$ is the large number of possible
displacement operators. We note that, since, on one hand,
${\mathbb Q}_{\mathcal V}$ and ${\mathcal M}^{\pm}$, and, on the
other other hand, ${\mathbb Q}_{\mathcal V}$ and ${\mathcal
B}^{\pm}$ are noncommuting operators, it becomes possible to define,
by any order of composition, the displacement operators. Adding now
${\mathbb Q}_{\mathcal V}^{-1}$ into the game, the set of
significant operators is even more large (for instance, note that
$[{\mathcal B}^{-}\widetilde{\mathbb Q}_{\mathcal V},
\widetilde{\mathbb Q}_{\mathcal V}^{-1}{\mathcal M}^{+}]  =
\mathbb{I}_{\mathcal V}$). The corresponding displacement operators
obviously generate different sets of NVCSs for a particular order
and by coupling ${\mathbb Q}_{\mathcal V}$ or ${\mathbb Q}_{\mathcal
V}^{-1}$ to ${\mathcal M}^{\pm}$ or ${\mathcal B}^{\pm}$. To study
all  these NVCSs is of course an interesting issue that we will
post-pone to a forthcoming work consisting in the classification of
these families using different criteria (for instance exact solution
of their moment problem  and which of them are simultaneously
solvable, or by sharper analytical properties, etc.). Finally, one
remarks that, in deformed theory, an initial family of NVCSs
could have many solvable ``dual" counterparts.\\

\subsection{Matrix displacement operators}

The same notion of
deformed displacement operators also makes  sense into a matrix
theory. In this paragraph, we only focus on normal matrix NVCSs, the
case of quaternionic domain could be  easily inferred. Consider the
matrix annihilation operator as given by (\ref{matann}) and the operator (same
notations as in Section \ref{sect5} are used)
\begin{eqnarray}
 \mathbb{B}_{V}^{+}=
\sum_{n=0,\pm}^\infty\,  |n+1\rangle\,\langle n |\otimes \,
V \, \overline{G(\{n+1\})} \,V^\dag|\pm\rangle\,\langle \pm|,
\quad V \in U(2),
\end{eqnarray}
such that $\overline{G(\{n\})} ={\rm diag}(\overline{G_+(\{n\}}, \overline{G_-(\{n\}})$.
Then, the following algebra is satisfied
\begin{eqnarray}
[\mathbb{M}_{V}^{-},\mathbb{B}_{V}^{+}] = \mathbb{I}_{\mathcal V},
\end{eqnarray}
if and only if, at the matrix level
\begin{eqnarray}
&&K(\{1\}) \overline{G(\{1\})}= \mathbb{I}_2,\\
&&K(\{p+1\}) \overline{G(\{p+1\})} - K(\{p\}) \overline{G(\{p\})}= \mathbb{I}_2,\;\;\;
\forall p\geq 1.
\end{eqnarray}
These problems have the solutions
\begin{eqnarray}
\overline{G(\{0\})}= 0, \qquad \overline{G(\{p\})} = p\, (K(\{p\}))^{-1},
\;\;\forall p\geq 1.
\end{eqnarray}
Consequently, the matrix operator
\begin{eqnarray}
&&\fl \mathbb{D}_{f}=  \exp\, [ {\mathbb B}_V^{+} \cdot \mathfrak{Z}_f
-\mathfrak{Z}^+_{f}\cdot\,{\mathbb M}_V^{-} ],\\
&&\fl  \mathfrak{Z}^+_{f} := \,(\mathbb{Q}^{V}_{\mathcal V})^{-1}\mathfrak{Z}^\dag\, = \sum_{n=0,\pm}^\infty\, |n\rangle\,\langle n|\otimes V\,\mathfrak{Z}^\dag_{\rm diag} (h_f(n))^{-1}\,  V^\dag|\pm\rangle\,\langle \pm|,
\end{eqnarray}
defines a displacement operator for normal matrix NVCSs.
Hence, we have
\begin{eqnarray}
 |\mathfrak{Z};\tau_\pm;\pm\rangle
=\mathbb{D}_{f} \left[ \exp\left[\frac{1}{2}\mathfrak{Z}^\dag\mathfrak{Z}\right]
N(\mathfrak{Z})\, V\, \exp[-i\omega_0 \tau e_0]V^\dag \,\;|0,\pm\rangle \right],
\label{eq:vcs122}
\end{eqnarray}
where $N(\mathfrak{Z})$ is the normalization factor given by
(\ref{eq:norma}) and the phase factor is introduced to maintain the
theory stable under the time evolution along the lines of Subsection
\ref{subsect41}. We should comment that, one should more
rigorously write the exponent of the exponential factor as written
in (\ref{eq:vcs122}) as follows
\begin{eqnarray}
\mathfrak{Z}^\dag\mathfrak{Z} = \sum_{n,\pm}^\infty
|n \rangle \langle n| \otimes \mathfrak{Z}^\dag\mathfrak{Z}
|\pm \rangle \langle \pm |  .
\end{eqnarray}
The duals of matrix displacement operators also have a well defined
sense. Indeed, defining $\mathbb{B}_{V}^-
=(\mathbb{B}_{V}^+)^{\dag}$, the algebra
\begin{equation}
[\mathbb{B}_{V}^{-},\mathbb{M}_{V}^{+}]
= \mathbb{I}_{\mathcal V},
\end{equation}
 is trivially satisfied. We introduce the deformed dual operator
\begin{eqnarray}
&& \mathbb{D}'_{f}=
\exp[{\mathbb M}^{+}\cdot\mathfrak{Z}_f
-\mathfrak{Z}^+_{f}\cdot\,{\mathbb B}^{-} ],
\end{eqnarray}
and deduce dual matrix NVCSs as
\begin{eqnarray}
 |\mathfrak{Z};\tau_\pm;\pm\rangle'
 &=&\mathbb{D}'_{f} \left[ \exp\left[\frac{1}{2}\mathfrak{Z}^+\mathfrak{Z}\right]
N'(\mathfrak{Z}) \exp[+i\omega_0\tau e_{0}] \, \;|0,\pm\rangle \right]\cr
&=& N'(\mathfrak{Z}) \sum_{n=0}^\infty  |n\rangle \otimes\,V\, \frac{R'^0(n)}{n!}\,
\exp[+i\omega_0\tau e_{n}] \, \mathfrak{Z}_{\rm diag}^n \, V^\dag\, |\pm\rangle,
\label{eq:vcs1222}\\
&& N'(\mathfrak{Z}) = \sum_{n=0}^{\infty}\,
\left[\frac{|z|^{2n}}{(n!)^2}(R'^0_{+}(n))^2
+ \frac{|w|^{2n}}{(n!)^2}(R'^0_-(n))^2\right],
\end{eqnarray}
where $R'^0(n) = K(\{n\})! (h^0_f(n-1)! h^0_f(0))$. A notable
feature of dual NVCSs is the change of sign of the phase factor in
(\ref{eq:vcs1222}). This is related to the fact that, if we keep the
same notation as in Subsection \ref{subsect41}, the state
$|\mathfrak{Z};\tau_\pm;\pm\rangle'$ evolves in the opposite time
direction relative to his proper time $\tau_\pm$.
Indeed, considering the time evolution operator $U_V(t)=
V\, \exp[-i \omega_0 t \mathbb{H} ]\, V^\dag$, we have the
Gazeau-Klauder temporal stability condition for dual states  which
is written as
\begin{eqnarray}
U_V(t) |\mathfrak{Z};\tau_\pm;\pm\rangle' =
 |\mathfrak{Z};\tau_\pm - t;\pm\rangle'.
 \end{eqnarray}
This shows that the dual state $|\mathfrak{Z};\tau_\pm;\pm\rangle'$
is a kind of ``proper time reversal state'' of its dual partner
$|\mathfrak{Z};\tau_\pm;\pm\rangle$. This can be also seen by
considering the following proper Schr\''odinger equations:
\begin{eqnarray}
&&(i\hbar) \partial_{\tau_\pm} |\mathfrak{Z};\tau_\pm;\pm\rangle = (i\hbar)(- i\tau_\pm \omega_0 \mathbb{H} ) |\mathfrak{Z};\tau_\pm;\pm\rangle, \\
&&(i\hbar) \partial_{\tau_\pm} |\mathfrak{Z};\tau_\pm;\pm\rangle' = (i\hbar)(+ i\tau_\pm \omega_0 \mathbb{H}) |\mathfrak{Z};\tau_\pm;\pm\rangle'.
 \end{eqnarray}
Again, a similar treatment as above
for finding solutions of the resolution of unity moment problems,
makes explicit dual matrix NVCSs. Still here, there
is a number of different displacement operators for NVCSs over
matrix domains.

\subsection{Deformed $T$ operators}
\label{subsect63}

{\bf Ordinary concept of $T$ operators}.
A type of operators allowing the mapping
between canonical and NCSs have been highlighted by
Ali {\it et al} \cite{rok2} and definitely used in \cite{rok}.
This mapping rests on the idea that one could transform CSs
into deformed ones via an operator called $T$ and into dual CSs via the inverse operator
$T^{-1}$, with $T T^{-1}= \mathbb{I}$, the identity onto the Hilbert
space considered. We have, using ordinary CS notations,
\begin{eqnarray}
|z\rangle \stackrel{T}{\longmapsto} |z\rangle_{f}, \qquad
|z\rangle\stackrel{T^{-1}}{\longmapsto}  |z\rangle'_f
\end{eqnarray}
For the simple instance of canonical CSs and NCSs, one gets
\begin{eqnarray}
\left[
|z\rangle = e^{-\frac{1}{2}|z|^2} \sum_{n =0}^{\infty} \frac{z^n}{\sqrt{n!}} |n\rangle\right]
\stackrel{T}{\longmapsto}
\left[ |z\rangle_{f} = N(|z|) \sum_{n =0}^{\infty} \frac{z^n}{\sqrt{x_n !}} |n\rangle
\right] \\
x_n ! := \prod_{k=1}^{n} x_k, \;\;\; x_0 ! := 1, \quad
N(|z|) = \left[ \sum_{n =0}^{\infty} \frac{|z|^{2n}}{x_n !} \right]^{-\frac{1}{2}},
\end{eqnarray}
with  $x_n$ a nonlinear function of $n$, i.e. the deformation function
on which one should impose the condition such that the norm series $N(|z|)$ converges
in a nonempty complex disc, and the operator
\begin{eqnarray}
T = N(|z|) e^{+\frac{1}{2}|z|^2} \sum_{n=0}^{\infty}\, \sqrt{\frac{n!}{x_n!}} |n\rangle\, \langle n|.
\label{t0}
\end{eqnarray}
A rapid verification shows that $T^{-1}$ is well defined and,
\begin{eqnarray}
\tilde T^{-1} =N'(|z|)\,N(|z|)\, e^{+\frac{1}{2}|z|^2}  T^{-1},\quad
N'(|z|) = \left[ \sum_{n =0}^{\infty} \frac{|z|^{2n}}{(n!)^2} (x_n !) \right]^{-\frac{1}{2}},
\label{tm1}
\end{eqnarray}
 maps $|z\rangle$ onto the dual state $|z\rangle'_f$.
Let us observe that, in foregoing studies, these $T$ and $T^{-1}$ do
not include normalization factors \cite{rok}. This could bring
ambiguities when one wants to map normalized CSs onto normalized
deformed one. As an answer of this issue, the definitions of $T$
(\ref{t0}) and $\tilde{T}^{-1}$ (\ref{tm1}) provide the correct
operators. Furthermore, the temporal stability axiom is not verified
by any of the  above-mentioned states. Let us find an equivalent
speech for canonical VCSs, NVCS and dual NVCSs and allowing the time
translations.

\noindent{\bf Matrix $T$ operators}. We only deal with the case of
normal matrix domains. One can easily deduce the results for
quaternionic and for $S^2$-NVCSs by a similar analysis. Consider
NVCSs over normal matrices, given $V\in U(2)$. Then, we define the
operators
\begin{eqnarray}
\fl T_f =N(|\mathfrak{Z}|)\, (N_0(|\mathfrak{Z}|) )^{-1}
\sum_{n=0,\pm}^{\infty} |n \rangle\, \langle n|\,\otimes \,V\, \sqrt{n!} \,R_0(n)\,e^{+i\omega_0\tau (e^{0}_{n}-e_{n})}  V^\dag\,|\pm \rangle\, \langle \pm|
\label{tops}\\
N_0(|\mathfrak{Z}|)  =  \left[ \sum_{n =0}^{\infty}
\left(\frac{|z|^{2n}}{n!} +  \frac{|w|^{2n}}{n!}\right) \right]^{-\frac{1}{2}},
\end{eqnarray}
where $e^{0}_{n}=\lim_{f(n)\to 1} e_n$ is the eigenenergies of the canonical model
and $N_0(|\mathfrak{Z}|)$ is the normalization factor of the canonical Gazeau-Klauder
VCSs
\begin{eqnarray}
|\mathfrak{Z}, \tau_\pm, \pm \rangle_0  =
N_0(|\mathfrak{Z}|) \sum_{n=0}^{\infty}|n\rangle \otimes\,V\, \frac{1}{\sqrt{n!}} e^{-i\omega_0\tau e^0_{n}}\;\mathfrak{Z}_{\rm diag}^n \, V^\dag |\pm\rangle,
\end{eqnarray}
which could be deduced from matrix NVCSs of Section \ref{sect5} by taking
deformation parameters limit $(\kappa,f(N))\to (1,1)$.
The phase of $T_f$ (\ref{tops}) will contribute to the temporal stability
of the resulting state.
A straightforward calculation gives
the correct mapping of (time stable and normalized) canonical VCSs onto
(time stable and normalized) NVCSs
\begin{eqnarray}
T_f |\mathfrak{Z}, \tau_\pm, \pm \rangle_0  = |\mathfrak{Z}, \tau_\pm, \pm \rangle.
\end{eqnarray}
Next, let us seek the operator mapping $|\mathfrak{Z}, \tau_\pm, \pm \rangle_0$ onto $|\mathfrak{Z}, \tau_\pm, \pm \rangle'$. Regarding the inverse
operator
\begin{eqnarray}
\fl T^{-1}_f =(N(|\mathfrak{Z}|))^{-1}\, N_0(|\mathfrak{Z}|)
\sum_{n=0,\pm}^{\infty} |n \rangle\, \langle n|\,\otimes \,V\, \frac{1}{\sqrt{n!}}(R_0(n))^{-1}\,e^{-i\omega_0\tau (e^{0}_{n}-e_{n})} |\pm \rangle\, \langle \pm|\, V^\dag,
\end{eqnarray}
clearly this does not furnish the correct answer if we would like to
respect all Gazeau-Klauder axioms. However, keeping in mind that
dual states are proper time reversal states of the original theory, the
operator ${\mathcal T}_f$ defined as
\begin{eqnarray}
{\mathcal T}_f =
(N(|\mathfrak{Z}|))^{2}\, (N_0(|\mathfrak{Z}|))^{-2}\,  T^{-1}_f
\end{eqnarray}
gives the answer
\begin{eqnarray}
{\mathcal T}_f \, |\mathfrak{Z}, - \tau_\pm, \pm \rangle_0 \,=\,
|\mathfrak{Z}, \tau_\pm, \pm \rangle',
\end{eqnarray}
indicating that canonical proper time reversal VCSs are mapped onto dual
NVCSs.

\section{A new class of $S^3$ NVCSs}
\label{sec4}

There is another class of exactly solvable NVCSs that we could
define on the Hilbert space ${\mathcal V}$ and being still continuous at $z=0$. It is worth
noticing that, although
the previous construction including the finite sequence of states
into one or another  tower well works, another alternative could  be also of interest.
Indeed, the considered finite sequence of $k$-initial states
of the Hilbert space can be viewed as a third part on its own,
not depending on the two towers, to which one can assign a new vector index
(here an angle parametrizing $S^3$).
The resulting NVCSs satisfy Gazeau-Klauder properties
and yield an exact solution of their moment problem associated
with the resolution of the identity.

The following states, that we shall refer to $S^3$ NVCSs, are
parameterized by the unit sphere $S^3$ vectors
labeled by the angles $(\Theta,\phi)$,
$\Theta=(\theta_1,\theta_2)$,
$\theta_i \in [0,\pi]$, $\phi\in [0,2\pi[$, and the real time
parameters $(\tau_*,\tau_\pm)$
\begin{eqnarray}
&& |z;(\tau_*,\tau_\pm);(\Theta,\phi)\rangle =
\ \ \
{\mathcal N}^*(|z|)\,\cos\theta_1\,
\sum_{q=0}^{k-1}\, \frac{z^q}{K^0_-(\{q\})!}\,
e^{-i\omega_0\tau_*\,e^-_{q}}\,|e^-_{q}\rangle \cr
&&
\ \ \ +{\mathcal N}^-(|z|)\,\sin\theta_1\cos\theta_2\,
\sum_{n=k}^\infty\, \frac{z^n}{K^0_-(\{n\})!}\,
e^{-i\omega_0\tau_-\,e^-_{n}}\,|e^-_{n}\rangle  \cr
&&\ \ \ + \ {\mathcal N}^+(|z|)\,e^{i\phi}\,\sin\theta_1\sin\theta_2\,
\sum_{n=0}^\infty \,
\frac{z^{n}}{K^0_+(\{n\})!}\,e^{-i\omega_0\tau_+\,e^+_{n}}\,|e^+_{n}\rangle,
\label{eq:fcoh2120}
\end{eqnarray}
where the norm series
\begin{eqnarray}
{\mathcal N}^*(|z|) = \left\{\sum_{q=0}^{k-1} \frac{|z|^{2q}}{(K^0_-(\{q\})!)^{2}}\,\right\}^{-\frac{1}{2}}, \\
{\mathcal N}^\pm(|z|) = \left\{\sum_{n=n_0^\pm}^{\infty} \frac{|z|^{2n}}{(K^0_\pm(\{n\})!)^{2}}\,\right\}^{-\frac{1}{2}},\;\;
 n_0^+= 0, \;\; n_0^-=k,
\end{eqnarray}
are such that $|z|\leq R$, $R=\min\left(R_+,R_-\right)$ being
minimum of the convergence radii $R_\pm = \lim_{n\to \infty}
(K^0_\pm(\{n\})$ of ${\mathcal N}^\pm(|z|)$. The positive functions
$K^0_\pm(\{n\})$ are, for the moment, still free.

We sketch the proof that the $S^3$ NVCSs (\ref{eq:fcoh2120}) are of
Gazeau-Klauder type.

\begin{enumerate}

\item[(i)] The normalization and continuity of labeling are clearly guaranteed
after a simple evaluation.

\item[(ii)] The time evolution of these states under the unitary operator
$U(t)= e^{-i \omega_0 t {\mathcal H}^{\rm red}}$ is such that
\begin{eqnarray}
U(t)|z;(\tau_*,\tau_\pm);(\Theta,\phi)\rangle =
|z;(\tau_*+t,\tau_\pm+t);(\Theta,\phi)\rangle,
\end{eqnarray}
so the total set of $S^3$ NVCSs is stable under time evolution.

\item[(iii)] Action angle variables are $(\{J_*,J_-,J_+\}, \{\tau_*,\tau_-,\tau_+\})$
such that
\begin{eqnarray}
J_*  = ({\mathcal N}^*(|z|))^2 (\cos\theta_1)^2
\sum _{q=0}^{k-1} \frac{|z|^{2q}}{(K^0_-(\{q\})!)^{2}} e^-_q, \\
J_- = ({\mathcal N}^-(|z|))^2 (\sin\theta_1\cos\theta_2)^2\,
\sum_{n=k}^{\infty} \frac{|z|^{2n}}{(K^0_-(\{n\})!)^{2}}e^-_n,\\
J_+ = ({\mathcal N}^+(|z|))^2 (\sin\theta_1\sin\theta_2)^2\,
\sum_{n=k}^{\infty} \frac{|z|^{2n}}{(K^0_+(\{n\})!)^{2}}e^+_n.
\end{eqnarray}
\item[(iv)] The resolution of the identity can be written as
\begin{eqnarray}
&& \mathbb{I}_{\mathcal V} = \sum_{n=0,\pm}^\infty\,|e^\pm_{n}\rangle\,\langle e^\pm_{n}|
\\
&&=\int_{D_R\times S^3}\,d\mu(z;\Theta,\phi)\,
|z;(\tau_*,\tau_\pm);(\Theta,\phi)\rangle\,\langle z;(\tau_*,\tau_\pm);(\Theta,\phi)|,
\label{eq:resolu}
\end{eqnarray}
\end{enumerate}
where the $S^3$ measure $d\mu(z;\Theta,\phi)$ has the parametrization
\begin{eqnarray}
&&  d\mu(z;\theta,\phi)=\,\sin\theta_1 \sin \theta_2\,d^2z\,d\theta_1\, d\theta_2\, d\phi\,
\left\{ W^+(|z|)\sum_{n=0}^\infty|e^+_{n}\rangle\langle e^+_{n}|\,+ \right. \cr
 && \ \ \ \ \ \ \left.
W^-(|z|)\sum_{n=k}^\infty|e^-_{n}\rangle\langle e^-_{n}|
+W^*(|z|)\sum_{n=0}^{k-1}|e^-_{n}\rangle\langle e^-_{n}| \right\},
\label{eq:wei}
\end{eqnarray}
here $W^*(|z|)$ and $W^\pm(|z|)$ are real weight functions, which
are yet unknown.

After substitution in (\ref{eq:resolu}), with again $z=r\,\exp(i\varphi)$ and a measure
in radial sector chosen as $d^2z=r\,dr \,d\varphi$ with $r\in[0,R)$ and $\varphi\in[0,2\pi[$,
 one comes to the moment problems
\begin{eqnarray}
&&0\leq n \leq k-1,\quad\;\;\int_0^{R^2} du\,u^n\,h_*(u)= \,(K^0_-(\{n\})!)^2,
\label{eq:sjes0}\\
&& n\geq n^\pm_0 ,\qquad\;\;\int_0^{R^2} du\,u^n\,h_\pm(u)= \,(K^0_\pm(\{n\})!)^2,
\label{eq:sjes}
\end{eqnarray}
where $u=r^2$ and the moment functions $h_*(r^2)$ and $h_\pm(r^2)$ are such that
\begin{eqnarray}
&& h_*(r^2) =\frac{8\pi^2}{3} |{\mathcal N}^*(r)|^2\,W^*(r),
\label{eq:hfunct0}\\
&& h_-(r^2)=\frac{16\pi^2}{9}\,|{\mathcal N}^-(r)|^2\,W^-(r),\\
&&
h_+(r^2)=\frac{32\pi^2}{9}\,|{\mathcal N}^+(r)|^2\,W^+(r).
\label{eq:hfunct}
\end{eqnarray}
In order to solve the problems (\ref{eq:sjes0})-(\ref{eq:hfunct}),
we can set  some constraints onto the free deformation function
$K^0_\pm(\{n\})$.  Let us observe some solutions in the undeformed
situation. We then map $(\kappa, f(N))\to (1,1)$, and set the
so-called action-identity constraint \cite{kl2} defined by the set
of relations
\begin{eqnarray}
J_*=\cos^2\theta_1 \left(|z|^2 + e^-_{0}\right),\label{eq:actid0}\\
J_-=\sin^2\theta_1 \cos^2 \theta_2 \left(|z|^2 + e^-_{k}\right),\\
J_+=\sin^2\theta_1\sin^2\theta_2 \left(|z|^2 + e^+_{0}\right).
\label{eq:actid}
\end{eqnarray}
We can infer from (\ref{eq:actid0})-(\ref{eq:actid}), at the
decoupled model limit $\lambda(N)\to 0$, the constraints
\begin{eqnarray}
&& K^0_-(n) = \sqrt{e^-_{n} - e^-_{0}}= \sqrt{(1+\epsilon)n},\qquad 0 \leq n\leq k-1, \\
&& K^0_-(n) = \sqrt{e^-_{n} - e^-_{k}}=\sqrt{(1+\epsilon)n},\qquad n\geq k,\\
&& K^0_+(n) = \sqrt{e^+_{n} - e^+_{0}}=\sqrt{(1+\epsilon)n},\qquad n\geq 0,
\end{eqnarray}
assuming a bounded from below energy spectrum, i.e. $e^\pm_{n} -
e^\pm_{0}\geq 0$. The subsequent norm series,
\begin{eqnarray}
&&|{\mathcal N}^*(r)|^{-2} = \sum_{q=0}^{k-1}
\frac{|z|^{2q}}{(1+\epsilon)^{q}q!},\\
&&|{\mathcal N}^-(r)|^{-2} = e^{\frac{r^2}{1+\epsilon}} -  |{\mathcal N}^*(r)|^{-2},\\
&&|{\mathcal N}^+(r)|^{-2} = e^{\frac{r^2}{1+\epsilon}},
\end{eqnarray}
are of infinite radii of convergence.
The moment problems can be easily performed with
solutions $h^*(r)=h^\pm (r) =\exp [-r^2/(1+\epsilon)]/(1+\epsilon)$, from
which we can deduce the weights
\begin{eqnarray}
&& W^*(r) =  \frac{3e^{-\frac{r^2}{1+\epsilon}} }{8\pi^2(1+\epsilon)} |{\mathcal N}^*(r)|^{-2} \quad
\label{eq:weit0}\\
&&
W^-(r)=\frac{9}{16\pi^2(1+\epsilon)}(1- e^{-\frac{r^2}{1+\epsilon}} |{\mathcal N}^*(r)|^{-2} ),
\label{eq:weit1}\\
&&W^+(r)= \frac{9}{32\pi^2(1+\epsilon)},
\label{eq:weit2}
\end{eqnarray}
then indicating  a new class of NVCSs.

We can turn the discussion to the deformed case.
If we set $K^0_\pm(\{n\})= \sqrt{\{n\}}$, we end with the
moment problems
\begin{eqnarray}
&&0\leq n \leq k-1,\quad\;\;\int_0^{R^2} du\,u^n\,h_*(u)= \,\{n\}!\,,
\label{eq:sj0}\\
&& n\geq n^0_\pm ,\qquad\;\;\int_0^{R^2} du\,u^n\,h_\pm(u)= \{n\}!\,,
\label{eq:sjs}
\end{eqnarray}
with $n^0_-=k$, $n^0_+=0$, whose  solutions can be provided in terms
of $(p,q)$ deformations as previously performed.

\section{Conclusion}
\label{sec6}

The construction of  new  Gazeau-Klauder type NVCSs for  spin-orbit
Hamiltonians has been achieved in this work. We have extended the
action of the ladder operators  to the initial finite dimensional
set of states related to the multi-photon processes. We have also
succeeded  in
 finding  exact solutions to the resolution of the identity for
different sets of NVCSs. Besides, we have addressed the issues of
different displacement and $T$ operators which  generate the variety
of states that we have found. Moreover, we have built a new class of
NVCSs parameterized by unit vectors of the $S^3$ sphere and proved
that the latter also generate an overcomplete set of VCSs. Finally,
it is worthy to emphasize  that Gazeau-Klauder axioms are nonempty
in the full deformation theory.

\ack The authors thank the referees for useful comments which allow
them to improve the paper. This work was supported under a grant of
the  National Research Foundation of South Africa and by the ICTP
through the OEA-ICMPA-Prj-15. The ICMPA is in partnership with the
Daniel Iagolnitzer Foundation (DIF), France.

\section*{Appendix}
 \renewcommand{\theequation}{A.\arabic{equation}}
\setcounter{equation}{0}

This appendix lists useful identities on $(p,q)$-deformed exponential functions.
We use the notations and convention of \cite{BGH2} with
$(p,q)$-shifted products and factorials defined as,
for any real parameters $a$, $b$ and $\alpha$
such that $a\neq 0$, $p>1$, $0<q<1$ and $pq < 1$,
\begin{eqnarray}
\fl [a,b;p,q]_{0}=1,\quad
[a,b;p,q]_\alpha=\frac{[a,b;p,q]_\infty}{[ap^\alpha,bq^\alpha;p,q]_\infty},\quad
[a,b;p,q]_\infty= \prod_{n=0}^\infty\left(\frac{1}{ap^n}-bq^n \right).
\end{eqnarray}
Given new parameters $(z,\mu,\nu) \in \mathbb{C}\times\mathbb{R}\times\mathbb{R}$,
the usual exponential function $e^{z}$, $z\in \mathbb{C}$,
can be extended to the generalized $(\mu,\nu,p,q)$-exponential as follows
\begin{eqnarray}
\label{eq:pqex}
{\mathcal E}_{(p,q)}^{(\mu,\nu)}(z)= \sum_{n=0}^{\infty}
\left( \frac{q^{\mu}}{p^{\nu}} \right)^{n^{2}}
\frac{z^{n}}{[p,q;p,q]_{n}},
\end{eqnarray}
provided $q^{2\mu}p^{1-2\nu}\le 1$.
The exponential function is recovered after rescaling
$z \to z(p^{-1}-q)$, for example, and taking the limit
$\lim_{(p,q)\to (1,1)}{\mathcal E}_{(p,q)}^{\mu,\nu}(z(p^{-1}-q))=e^{z}$.
Through the reduction $\mu=0$ and $\nu=1/2$,
(\ref{eq:pqex}) generates another $(p,q)$-exponential as
\begin{eqnarray}
e_{(p,q)}(z) =\sum_{n=0}^{\infty}
\frac{1}{p^{n^{2}/2}}\frac{z^{n}}{[p,q;p,q]_{n}},\qquad |z|<p^{-1/2}.
\label{eq:epqex}
\end{eqnarray}
The next identity stands for the $(p,q)$-analogue Euler Gamma function,
say the $(p,q)$-analogue of Ramanujan integral, for any $n\in\mathbb{N}$,
\begin{eqnarray}
\int_{0}^{\infty} dt\,t^n\,e_{(p,q)}\left(-\lambda_0 p^{-1/2}t\right)
= \frac{[p,q;p,q]_{n}}{\lambda^{n+1}_0\,q^{n(n+1)/2}} \log\left(\frac{1}{pq}\right).
\label{eq:App-Ramanujan2}
\end{eqnarray}

\end{document}